\documentstyle[12pt,aaspp4,flushrt]{article}

\slugcomment{Submitted to \aj}

\newcommand{\annm}{Ann. Numer. Math.}
\newcommand{\cmech}{Celest. Mech.}
\newcommand{\cmda}{\cmech\ Dyn. Ast.}
\newcommand{\icar}{Icarus}
\newcommand{\mbf}[1]{\mbox{\boldmath ${#1}$}}
\newcommand{\beq}{\begin{equation}}
\newcommand{\eeq}{\end{equation}}
\newcommand{\ba}{\begin{eqnarray}}
\newcommand{\ea}{\end{eqnarray}}
\newcommand{\cE}{{\cal E}}
\newcommand{\cM}{{\cal M}}
\newcommand{\emin}{e_{\rm min}}
\newcommand{\emax}{e_{\rm max}}
\newcommand{\jmax}{j_{\rm max}}
\newcommand{\Smax}{S_{\rm max}}
\newcommand{\Scrit}{S_{\rm crit}}
\newcommand{\kcrit}{k_{\rm crit}}
\newcommand{\torb}{t_{\rm orb}}
\newcommand{\tlib}{t_{\rm lib}}
\newcommand{\tdrift}{t_{\rm drift}}
\newcommand{\dtres}{\Delta t_{\rm res}}
\newcommand{\dt}{\Delta t}

\begin{document}
\null\vskip -0.5truein

\title{Dynamical Chaos in the Wisdom-Holman Integrator:\\ Origins and Solutions}

\author{\it Kevin P. Rauch}
\affil{Dept. of Astronomy, Univ. of Maryland, College Park, MD 20742-2421}
%\and
\author{\it Matthew Holman}
\affil{SAO, Mail Stop 18, 60 Garden St., Cambridge, MA 02138}

\begin{abstract}
We examine the non-linear stability of the Wisdom-Holman (WH)
symplectic mapping applied to the integration of perturbed, highly eccentric
($e\gtrsim 0.9$) two-body orbits.  We find that
the method is unstable and introduces artificial chaos into
the computed trajectories for this class of problems, {\it unless} the
step size is chosen small enough to always resolve periapse, in
which case the method is generically stable. This `radial orbit instability'
persists even for weakly perturbed systems.
Using the Stark problem as a fiducial test case,
we investigate the dynamical origin of this instability and
show that the numerical chaos results from the overlap of step size
resonances (cf. Wisdom \& Holman 1992); interestingly, for the Stark problem 
many of these resonances appear to be absolutely stable.

We similarly examine the robustness of several alternative integration methods: 
a regularized version of the WH mapping suggested by Mikkola (1997);
the potential-splitting (PS) method of Lee et al. (1997); and
two methods incorporating approximations based on Stark motion instead of
Kepler motion (cf. \cite{newet97}).
The two fixed point problem and a related, more general problem are used to
comparatively test the various methods for several types of
motion.  Among the tested algorithms,
the regularized WH mapping is clearly the most efficient
and stable method of integrating eccentric, nearly-Keplerian orbits 
in the absence of close encounters.
For test particles subject to both high eccentricities and very
close encounters, we find an enhanced version of the PS
method---incorporating time regularization, force-center switching, and an
improved kernel function---to be both economical and highly versatile. We
conclude that Stark-based methods are of marginal utility in $N$-body type
integrations. Additional
implications for the symplectic integration of $N$-body systems are discussed.
\end{abstract}


\section{Introduction}
\label{sec_intro}

Symplectic integration schemes have become increasingly popular tools for the 
numerical study of dynamical systems,
a result of their often high efficiency
as well as their typical long-term stability
(see, e.g., \cite{marps96} and the many references within).
The Wisdom-Holman (WH) symplectic mapping in particular
(\cite{wish91}; cf. \cite{kinyn91}) has been widely used 
in the context of Solar System dynamics.
However, the fact that this and other symplectic methods
are, by construction, ``finely tuned'' can make them susceptible to
performance-degrading ailments (much as high-order methods offer little
benefit if the motion is not sufficiently smooth), and the stability of these
methods for arbitrary systems and initial conditions is not completely
understood.  It would be prudent, therefore, to exercise caution when applying
such schemes to systems entering previously unexplored dynamical states, and
to ensure that adequate preliminary testing is undertaken regardless of the
method's stability in previously considered problems.
Recent galactic dynamics simulations
by Rauch \& Ingalls (1997), for example---which used the WH mapping---uncovered
evidence of an instability in the method when applied to a
particular class of problems: the integration of highly elliptical,
nearly-Keplerian orbits in which the timestep is taken small enough to smoothly
resolve the perturbation forces, but not so tiny as to explicitly resolve
pericenter. Since particle motion in these simulations was extremely close to
Keplerian near pericenter, and since the mapping itself is exact for Keplerian
motion, there is no {\it a priori} reason why the method should have performed
as poorly and unstably as was found.

Recently several variations of the WH mapping have been proposed
which aim to extend the range of applicability of the original method.
The regularized WH mappings investigated by Mikkola (1997), for instance,
appear promising in the context of elliptical motion. The potential-splitting
(PS) method of Lee, Duncan, \& Levison (1997) allows symplectic
integration of close
encounters between massive bodies by adding a multiple-timestep algorithm
similar to that of Skeel \& Biesiadecki (1994).
Unfortunately both of these schemes have limitations of their own;
the former is unable to resolve close encounters, while the latter approach
(like the original mapping) appears to be unstable 
when orbits are eccentric (cf. \cite{dunll97}).

In this paper, we use a series of test problems
based on perturbed two-body motion
to analyze the stability of the WH mapping and several of its variants.
In particular, we examine the reliability of the methods for test
particles whose motion is either highly eccentric or subject to close
encounters with the perturbers (or both).
The plan of the paper is as follows. In the following section, the performance
of the WH mapping at high eccentricities is investigated using the
Stark problem (e.g., \cite{dan94}; \cite{kir71})---for which the range of
orbital eccentricities is easily controlled, and no close encounters occur---as
the fiducial test case. The instability found in the integrated motion 
is then explained using complementary analytic and geometric arguments.
In \S~\ref{sec_modwh}, modified forms of the original mapping are described;
similarly, in \S~\ref{sec_sint} integrators based on Stark motion
instead of Kepler motion are considered.
Section~\ref{sec_compsim} uses the two fixed point problem (e.g., \cite{par65})
as well as a more general test problem (drawn from the area of galactic
dynamics) to conduct a comparative performance analysis of the various
algorithms.  Both the Stark and two fixed point problems are fully integrable
and analytically soluble in terms of elliptic functions and integrals,
allowing a detailed assessment of the accuracy of the numerical results to be
made. Concluding discussion is given in \S~\ref{sec_discuss}.

\section{The Stark Problem}
\label{sec_stark}

\subsection{Orbital Motion}
\label{sec_sorb}

The Stark problem represents the motion of a test particle about a fixed
Newtonian force center (i.e., the Kepler problem) subject in addition to a
uniform
force of constant magnitude and direction. The Hamiltonian (per unit mass) for
the problem is given by
\beq \label{eq_Hstark}
H={\mbf{p}^2\over 2}-{GM\over r}-\mbf{S}\cdot\mbf{x},
\eeq
where $\mbf{x}$ is the Cartesian position, \mbf{p} is its conjugate 
momentum (in this case, the physical velocity), $r=|\mbf{x}|$, $M$
is the mass of the central object, and the constant vector $\mbf{S}$ (the
`Stark vector') embodies the uniform field.
Physical analogies include: the (classical) orbit of an electron about a fixed
nucleus immersed in a uniform electric field; the trajectory about its parent
of an artificial satellite with a continuously firing thruster; and the motion
of a dust particle around a comet nucleus taking into account the radiation
pressure of sunlight (cf. \cite{ham92}; \cite{mig82}).
There are three conserved quantities (and hence the motion is regular): the
energy, $E$, the angular momentum component along the Stark vector,
$\mbf{L}\cdot\mbf{\hat S}$, and a third integral, $\alpha$ (say), which 
arises as a `separation constant' in the analytic solution of the problem.

\begin{figure}
\plotone{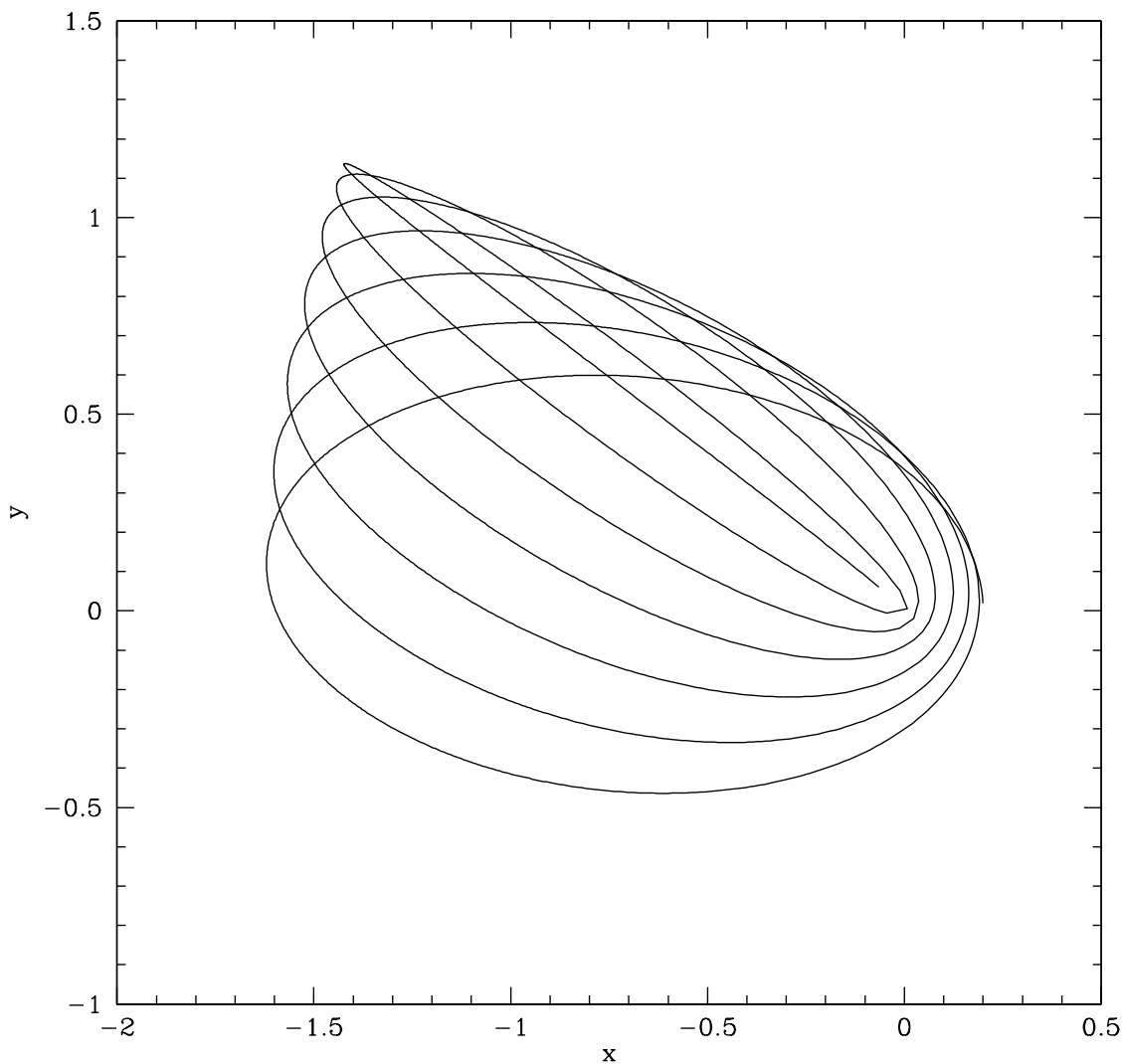}

\caption{
An example of bound, nearly-Keplerian Stark motion; the Stark vector is
directed along the $+{\hat x}$ direction. The initially prograde,
moderate eccentricity ($e_0=0.8$) orbit begins with its line of apsides along
the $x$-axis. The orbit subsequently suffers a clockwise precession and
becomes increasingly radial. Upon reaching $e=1$, the orbit switches to
retrograde motion and counterclockwise precession (not shown); the orbit
eventually returns to prograde circulation in the lower left quadrant, and the
cycle continues.
\label{fig_sorb}}
\end{figure}

Qualitatively speaking, motion in the Stark problem is bound whenever
$E<0$ and $|\mbf{S}|=S\lesssim \Scrit(E)=E^2/(GM)$, and
it is nearly-Keplerian when $S\ll \Scrit$.  In this latter
case the orbit consists of a precessing ellipse of varying eccentricity
and inclination. In the 2-D case ($z=p_z=S_z=0$, say) the maximum
eccentricity reached, $\emax$,
is always unity, and thus these orbits momentarily become radial (at
which time the circulation of the orbit changes from prograde to retrograde,
or vice versa); the minimum eccentricity, $\emin$, normally
occurs where the line of apsides is
parallel to the Stark vector and can lie anywhere between 0 and 1.
%{\it Special case: $e<1$ and line of apsides perpendicular to Stark vector...}
In the 3-D case, conservation of the angular momentum component forces
$\emax<1$, and now the inclination also varies between minimum and maximum
values.  In all cases, $\emin$ and $\emax$ are constants of the motion;
this conveniently allows fine control over the range of eccentricities
encountered during testing, regardless of the length of the integration.
An example of nearly-Keplerian ($S=0.12\Scrit$),
2-D motion is given in Figure~\ref{fig_sorb}.

\subsection{Behavior of the Wisdom-Holman Mapping}
\label{sec_whmap}

Using the WH mapping as a numerical integrator for the Stark problem is
equivalent (within round-off error) to replacing the
Hamiltonian~(\ref{eq_Hstark}) with a ``nearby'' mapping Hamiltonian, and
solving 
the resulting equations of motion exactly (see \cite{wish91}).
In the present case, a second order mapping Hamiltonian corresponding to
equation~(\ref{eq_Hstark}) is
\beq \label{eq_Hsmap}
H_{\rm map}=\left({\mbf{p}^2\over 2}-{GM\over r}\right)-
2\pi\delta_{2\pi}(\Omega t-\pi) (\mbf{S}\cdot\mbf{x}),
\eeq
where $\delta_{2\pi}(x)$ is a periodic delta function (with period $2\pi$),
$\Omega=2\pi/\Delta t$ is the mapping frequency, and $\Delta t$ is the
associated integration step size. (One physical realization of this Hamiltonian
is an orbiting satellite performing periodic, short-duration burns of
its engine.) The mapping Hamiltonian differs from the
original by a series of high frequency ($\Omega$ and higher harmonic) terms.
In general, the long-term evolution under $H_{\rm map}$ can be expected to
remain ``close'' to the true evolution as long as the mapping frequency
exceeds all fundamental dynamical frequencies in the problem;
this is known as the {\it averaging principle}. The second order integration
step corresponding to $H_{\rm map}$ is
\beq
\label{eq_sint_step}
I(\mbf{x}, \mbf{p}; \Delta t, \mbf{S})=D(\mbf{x}, \mbf{p},
\Delta t/2)\, K(\mbf{S}, \Delta t)\, D(\mbf{x}, \mbf{p}, \Delta t/2),
\eeq
where $D$ represents a drift along an unperturbed Keplerian orbit and $K$
represents a momentum kick due to the perturbation $\mbf{S}$ ($\mbf{x}$ is
left unchanged by the map $K$). We remark that the Kepler step $D$ is most
efficiently computed using the Gauss $f$ and $g$ functions
(e.g., \cite{dan92}).

\begin{figure}
\plotone{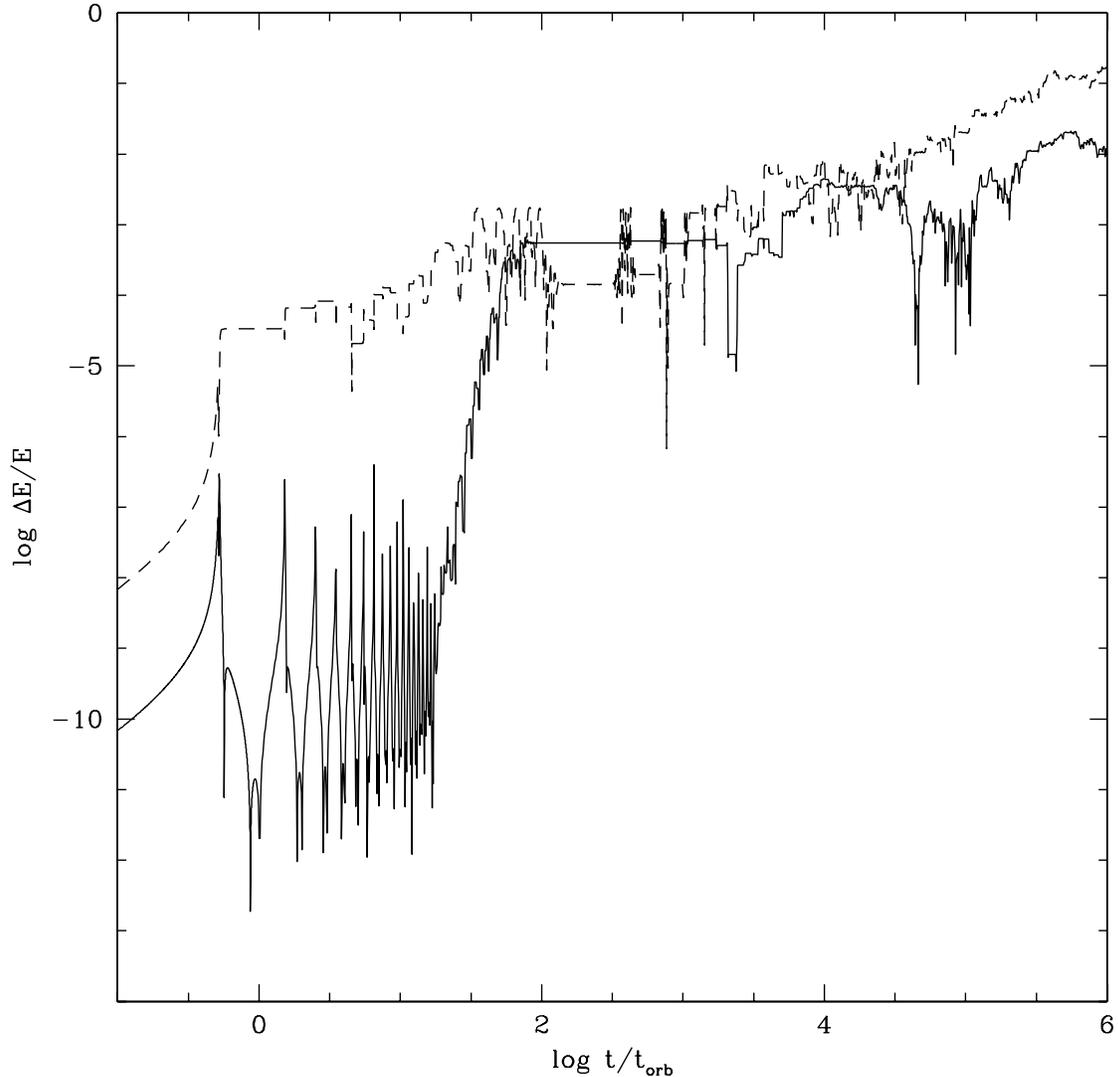}

\caption{
The fractional energy error over time committed in the integration of an
$S=4\times 10^{-3} \Scrit$, $e_0=0.9$, 2-D 
Stark orbit using the Wisdom-Holman method; the solid curve is for
$\dt=10^{-3}\torb$, the dashed for $\dt=10^{-2}\torb$.
The initially oscillatory behavior of the solid curve fails when $e$
increases to where the step size no longer resolves periapse; at late times,
both orbits random walk in energy and eventually go unbound. Note that
although the initial separation of the curves is just what would be expected
from a second order integrator, the separation is not even qualitatively
maintained over the long run.
\label{fig_rwalk1}}
\end{figure}

A typical example of the evolution under $H_{\rm map}$ is shown in
Figure~\ref{fig_rwalk1}, which plots the fractional energy error committed 
for a pair of 2-D integrations using 100 (dashed curve) and 1000 (solid curve)
points per orbit; the beginning
orbit, similar to Fig.~\ref{fig_sorb}, had an initial eccentricity $e_0=0.9$
and a Stark perturbation $S=4\times 10^{-3}\Scrit$ directed at a $45^\circ$
angle from the initial line of apsides. 
In contrast to the bounded, oscillatory energy errors typically
observed in symplectic integrations, in this case the errors undergo a random
walk towards unity, even using 1000 points per orbit.
Eventually, in fact, these numerical orbits become unbound and escape to
infinity---a
{\it qualitatively} incorrect result for the long-term motion of the particle.
This is suggestive (though not conclusive) evidence for numerical chaos,
which if present must arise from the integrator itself since the analytic
motion is fully integrable.
Although this result does not, strictly speaking, violate the averaging
principle (because the mapping frequency does not exceed the pericentric
passage frequency when the orbit is nearly radial), it is nonetheless
surprising that the orbits are so unstable under such a modest
perturbation---particularly
since the relative perturbation strength at periapse is even smaller.

\begin{figure}
\plotone{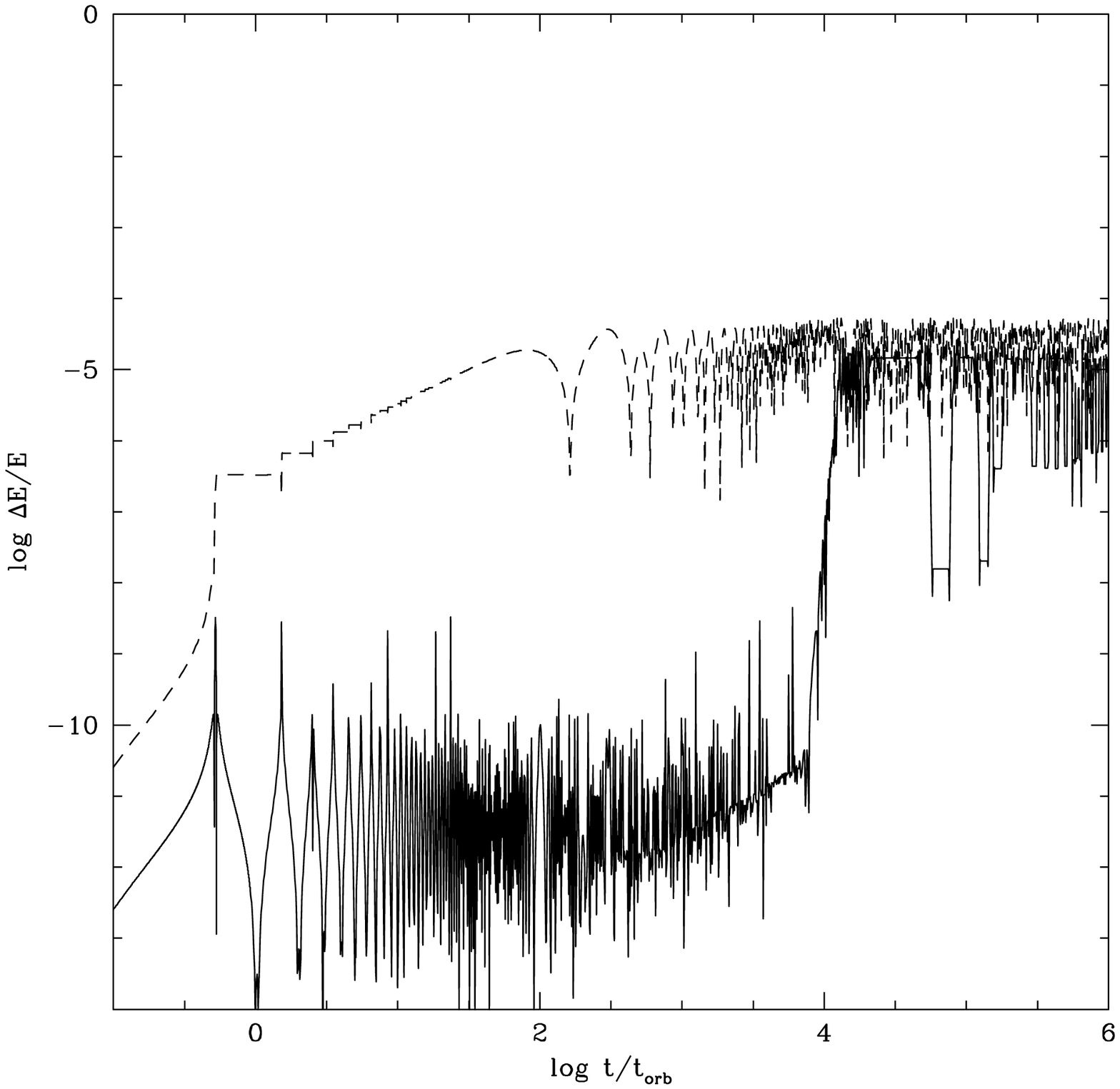}

\caption{
As Figure~\ref{fig_rwalk1}, but for $S=4\times 10^{-5} \Scrit$ (with $\mbf{S}$
parallel to the initial line of apsides). Note the bounded energy error at late
times.
\label{fig_rwalk2}}
\end{figure}

A less typical (but not infrequent) example of the evolution under $H_{\rm
map}$ is shown in Figure~\ref{fig_rwalk2}, which is analogous to
Figure~\ref{fig_rwalk1} except that the Stark vector
$S=4\times 10^{-5} \Scrit$ and is parallel to the initial semi-major axis.
As before, the 1000-point-per-orbit integration initially shows bounded,
oscillatory energy error which degrades significantly once periapse in no
longer resolved.  In this case, however, both integrations
appear to possess bounded
energy errors at late times, albeit at a much higher level than is exhibited
initially. This long-term stability generally persists indefinitely;
we have let particular simulations run for $\sim 10^9$ orbital periods with
no apparent change in the error bound.

\begin{figure}
\plotone{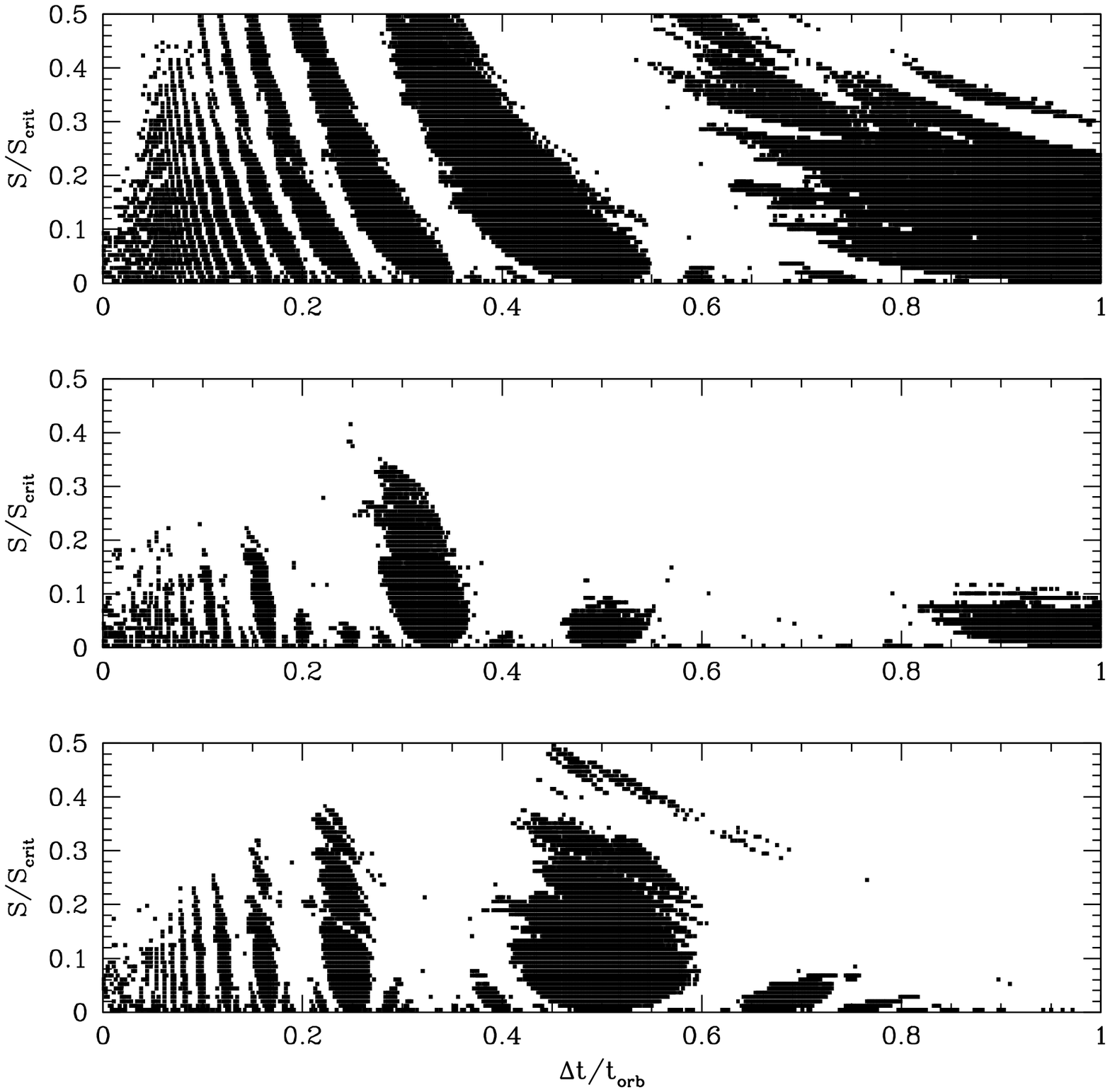}

\caption{
Islands of stability in the $\dt-S$ plane (regions smaller than a few pixels
are noise). Each panel corresponds to identical
initial conditions except for the starting orbital phase. 
The initial mean anomalies, from top to bottom, are $\cM_0=0, 2\pi/3, \pi$.
The islands all center on rational fractions of $\Delta t/\torb$, but
their shapes depend strongly on $\cM_0$. See \S~\ref{sec_geom}.
\label{fig_islands}}
\end{figure}

A strongly suggestive illustration of the underlying
distinction between the two types of
behavior is show in Figure~\ref{fig_islands}. The figure contains three
panels, each showing the regions of stability in the $\Delta t-S$ plane for
identical initial conditions aside from a change in orbital phase;
in each case $e_0=0.9$ and
the Stark vector is parallel to the initial line of apsides.
The initial mean anomalies of the orbits, from top to bottom,
are $\cM_0=0,\, 2\pi/3,\, \pi$. The `islands' of stability present in the panels
display a clear pattern: they are all centered on step sizes which are 
a rational fraction of the orbital period, and are most prominent at step sizes
corresponding to an integral number of points per orbit.
This is clear evidence that the {\it stable} behavior is the result of
step size resonances, and that otherwise the mapping is generically {\it
unstable}; a formal analytic analysis supporting
this assertion will be presented in the following section.
Note also how the
size of any individual island varies with the initial orbital phase---its area
reaches a maximum when one of the integration steps regularly lands near
periapse. The 2:1 ($\Delta t/\torb=1/2$) resonance island, for instance is
large in the $\cM_0=0$ and $\cM_0=\pi$ panels (for which every other step---the
odd or even ones, respectively---falls near pericenter), but small in the
$\cM_0=2\pi/3$ panel (where all steps fall rather far from pericenter). This
points to another necessary condition for stability; namely, stability does
not require that periapse be {\it resolved} (which is not even possible in two
dimensions, since $\emax=1$), but merely that it be {\it sampled}. We explain
these observations in detail in \S~\ref{sec_numstab}.

\begin{figure}
\plotone{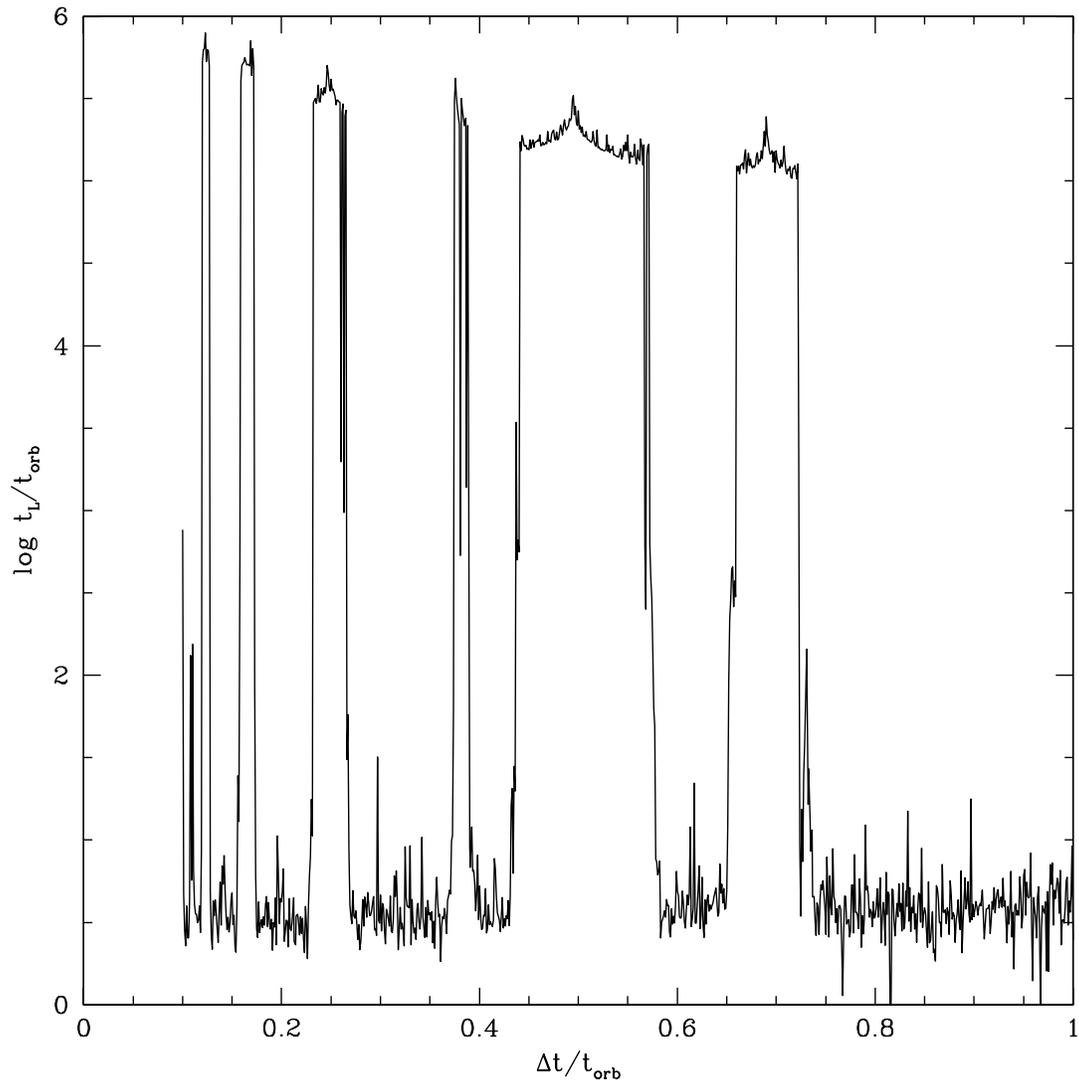}

\caption{
The Lyapunov times $t_{\rm L}$ across a horizontal slice (at $S/\Scrit=0.04$)
of the $\cM_0=\pi$ panel
of Figure~\ref{fig_islands}; values of $t_{\rm L}\gtrsim 10^5\torb$ should be
regarded as lower limits due to the finite length of the integration.
The location of the stable islands is clearly visible, indicating that these
are regions of true dynamical regularity (and not bounded chaos).
\label{fig_lyapunov}}
\end{figure}

Although true dynamical chaos is the most natural explanation for the
unstable behavior present in Figure~\ref{fig_rwalk1}, none of the evidence
presented so far proves that the ostensibly stable trajectories in
Figure~\ref{fig_rwalk2} are in fact dynamically stable---i.e., regular.
To test this, we computed the Lyapunov times (e.g., \cite{bengs76})
for a series of trajectories corresponding to a horizontal slice
(at $S=0.04\Scrit$) across the bottom panel of
Figure~\ref{fig_islands}. The result is shown in Figure~\ref{fig_lyapunov};
note that all values of $t_{\rm L}\gtrsim 10^5\torb$ should be regarded as
lower limits due to the finite length of the integration.
The figure provides convincing evidence that the `stable' islands are, in fact,
dynamically stable, and not an example of bounded chaos; off the islands, by
contrast, the motion is strongly chaotic. Examination 
of sample surfaces of section confirmed this result, and also verified that
the integrated motion inside the stable islands remains close (in terms of
explored phase space volume) to the analytic trajectory for arbitrarily long
times.

The preceding results do not qualitatively change when the general
three-dimensional problem is considered. The stable islands are still present,
although they shrink---and eventually disappear
(except the one for which $\dt$ resolves $\emax$)---as the Stark vector is
rotated from the $x$-axis (the initial line of apsides) to the $z$-axis
(perpendicular to the initial orbital plane); however, the same phenomenon also
occurs in the 2-D case when $\mbf{S}$ is rotated from the $x$-axis to the
$y$-axis. In both cases the disappearance of the islands coincides with
the condition $\emin\sim 0$; the connection is explained in \S~\ref{sec_geom}.

Finally, a few words about numerics.  Because of the extreme range of orbital
eccentricities encountered, the Stark problem places
severe stress on the Kepler stepper (the drift operator in
equation~[\ref{eq_sint_step}]) used in the integration;
the kick operator, by contrast, is completely trivial.
In fact, some of the stepper routines at our disposal (originally written with
low to moderate eccentricities in mind) failed outright for nearly radial
orbits, particularly at large step sizes. Among those proving robust,
however---including one using extended precision arithmetic
throughout---the choice of stepper did not
alter the stability of the mapping aside from minor changes in the
precise boundaries between the stable and unstable
regions. Our testing indicates that straightforward implementation of the
universal variables formulation of the Kepler stepper, combined with
constrained solution of Kepler's Equation (i.e., one guaranteed to converge)
and renormalization of the final radius and/or velocity vector
(to explicitly enforce energy conservation), produces a nearly optimal routine
in terms of both efficiency and robustness.

\subsection{Non-Linear Stability Analysis}
\label{sec_numstab}

Analytic
examination of the non-linear stability of the WH method was carried out using
the resonance overlap criterion described in Wisdom \& Holman (1992).
In brief, the method involves expansion of the mapping Hamiltonian as a sum of
resonant terms, with only the term representing the step size resonance under
consideration being retained. Expansion of this Hamiltonian about the resonant
value of the canonical momentum then leads to expressions for the width and
libration frequency of that particular resonance. Overlap, and subsequent
instability, occurs when the allowable libration amplitudes (in the energy
oscillations, say) of adjacent resonances exceeds their separation. 

\subsubsection{Hamiltonian Development}
\label{sec_anstab}

To begin the analysis we rewrite the mapping Hamiltonian in
equation~(\ref{eq_Hsmap}) to explicitly show the interaction of the
step size dependent terms with the terms of the original Hamiltonian. 
Using the Fourier representation of the delta functions,
\beq
\delta_{2\pi}(\Omega t) = \frac{1}{2\pi}\sum_{i=-\infty}^\infty \cos(i
\Omega t),
\eeq
the first order mapping Hamiltonian can be written as
\beq
H_{map} = \left(\mbf{p}^2 - \frac{GM}{r}\right) - \sum_{i=-\infty}^{\infty}
\cos(i \Omega t) \mbf{S}\cdot\mbf{x}.
\eeq
The time step, or interval between delta functions, is
$\dt=2\pi/\Omega$.  To simplify the analysis we here consider the 2-D
case with the Stark vector parallel to the x-axis.  The quantity
$\mbf{S}\cdot\mbf{x}$ is then $S_x x = S_x r \cos(\theta) = S_x r
\cos(\omega_l + f)$.  Here $r$ is the ``heliocentric'' distance,
$\omega_l$ is the longitude of pericenter, and $f$ is the true
anomaly.  Thus, the Hamiltonian is 
\ba
H_{map} &=& \left(\mbf{p}^2 - \frac{GM}{r}\right) - \sum_{i=-\infty}^{\infty}
\cos(i \Omega t) S_x r \cos(\omega_l + f) \\
&=& \left(\mbf{p}^2 - \frac{GM}{r}\right) - \sum_{i=-\infty}^{\infty}
\cos(i \Omega t) S_x \left(r \cos f\cos \omega_l - r \sin f \sin \omega_l\right).\\
\ea
Next, we expand $r \cos f$ and $r\sin f $ in terms of Bessel functions,
assuming a Keplerian orbit.
\ba
\frac{r}{a} \cos f &=& \cos(E) - e = -\frac{3e}{2} + 
2 \sum_{k=1}^{\infty}\frac{1}{k} J_k^\prime(k e) \cos(k \cM),\\
\frac{r}{a} \sin f &=& \sqrt{1-e^2}\sin(E) = \sqrt{1-e^2} \sum_{k=1}^{\infty}\frac{1}{k} J_{k-1}(k e) \sin(k \cM),
\ea
where $a$ is the semi-major axis, $E$ is the eccentric anomaly, $e$ is
the eccentricity, $\cM$ is the mean anomaly, $k$ is an integer
sequence, and  $J_{k-1}(k e)$ and $J_k^\prime(k e)$ are a Bessel
functions and their derivatives (\cite{broc61}).  Thus, the
Hamiltonian becomes 
\ba
H_{map} &=& \left(\mbf{p}^2 - \frac{GM}{r}\right) \\
& & - a S_x \cos \omega_l \sum_{i=-\infty}^{\infty} 
\cos(i \Omega t) \left[ -\frac{3e}{2} + 2
\sum_{k=1}^{\infty}\frac{1}{k} J_k^\prime(k e) \cos(k \cM)\right] \\
& & + a S_x \sin \omega_l \sum_{i=-\infty}^{\infty} 
\cos(i \Omega t) \left[\sqrt{1-e^2}\sum_{k=1}^{\infty}\frac{1}{k} J_{k-1}(k e) \sin(k \cM)\right],
\ea
where we ignore for the moment that $e$ and $\cM$ are not appropriate
canonical variables. The trigonometric factors can be re-arranged and combined:
\ba
H_{map} &=& \left(\mbf{p}^2 - \frac{GM}{r}\right)\\\nonumber
& & + a S_x \cos \omega_l \sum_{i=-\infty}^{\infty} \cos(i \Omega t) \frac{3e}{2} \\\nonumber
& & - a S_x \cos \omega_l \sum_{i=0}^{\infty} \sum_{k=1}^{\infty}\frac{1}{k} J_k^\prime(k e) \cos(k \cM - i \Omega t)\\\nonumber
& & + a S_x \sin \omega_l \sum_{i=0}^{\infty} \sum_{k=1}^{\infty}\sqrt{1-e^2} \frac{1}{k} J_{k-1}(k e) \sin(k \cM - i \Omega t).
\ea
Next, we make two related simplifying assumptions.  The first is that
$S_x$ is sufficiently small that the timescale for changes in the
eccentricity $e$ and the longitude of pericenter $\omega_l$ is
much longer than the orbital period or other timescales in the
system.  Thus, $e$ and $\omega_l$ will be considered as constants.  By
this assumption the terms in the first summation are strictly time
dependent and will not contribute to the equations of motion resulting
from the Hamiltonian.  We thus ignore those terms.
The second assumption is that the eccentricity is sufficiently large
that the $\sqrt{1-e^2}$ factor is negligibly small.  Dropping
the  final summation, 
\ba
H_{map} &\approx& \left(\mbf{p}^2 - \frac{GM}{r}\right)\\\nonumber
& & - a S_x \cos \omega_l \sum_{i=0}^{\infty} \sum_{k=1}^{\infty}\frac{1}{k} J_k^\prime(k e) \cos(k \cM - i \Omega t).
\ea
Re-writing the Hamiltonian in canonical variables:
\beq
H_{map} \approx - \frac{(GM)^2}{2L^2} - a S_x \cos \omega_l \sum_{i=0}^{\infty}
\sum_{k=1}^{\infty}\frac{1}{k} J_k^\prime(k e) \cos[k (\lambda - \omega_l) - i \Omega t],
\eeq
where the momentum $L = \sqrt{GM a}$ is canonically conjugate to the mean longitude
$\lambda = \omega_l + \cM $. The arguments to the cosine terms are now
clearly the ``step size resonances'' in the
Hamiltonian.  Such a resonance occurs when one of the
terms of the form $k (\lambda -\omega_l) - i \Omega t$ is slowly
varying.  Given our assumption that $\omega_l$ is roughly constant, this
happens when $k \dot{\lambda} - i \Omega \approx 0 $, i.e.,
when the orbital period is rationally related to the step size. 

Next we make a canonical change of variables that focuses on one of
these terms.  For this we use the mixed-variable generating function,
\beq
F_2 = (\lambda - \omega_l - \frac{i}{k}\Omega t)\Sigma + \lambda \Lambda,
\eeq
which results in the following transformation:
\ba
\sigma  &=& \frac{\partial F_2}{\partial \Sigma} = (\lambda - \omega_l) - \frac{i}{k} \Omega t,\\\nonumber
\lambda^\prime &=& \frac{\partial F_2}{\partial \Lambda} = \lambda,\\\nonumber
L  &=& \frac{\partial F_2}{\partial \lambda} = \Lambda + \Sigma.
\ea
The new Hamiltonian is
\ba
H^\prime &=& H + \frac{\partial F_2}{\partial t}\\
&=& -\frac{(GM)^2}{2(\Lambda + \Sigma)^2} - a S_x \cos \omega_l \frac{1}{k}
J_k^\prime(ke)\cos(k\sigma) - \frac{i}{k} \Omega \Sigma,
\ea
where only the resonant term has been retained.

Since we expect the canonical momentum to be constrained to a narrow
range of values at resonance, $\Sigma$ should vary little
from the resonant value.  Consider only the momentum terms in the
Hamiltonian:
\beq
H_0^\prime = -\frac{(GM)^2}{2(\Lambda + \Sigma)^2} - \frac{i}{k} \Omega \Sigma,
\eeq
which can be expanded about the resonant value of $\Sigma$ ($\Sigma^*$):
\beq
H_0^\prime = H_0^\prime|_{\Sigma^*} + \left.\frac{\partial
H_0^\prime}{\partial \Sigma}\right|_{\Sigma^*} (\Sigma - \Sigma^*) + \frac{1}{2}\left .\frac{\partial^2
H_0^\prime}{\partial \Sigma^2}\right|_{\Sigma^*} (\Sigma - \Sigma^*)^2.
\eeq
The first term is a constant and can be ignored.  The second term
defines the resonance condition
\beq 
\left.\frac{\partial H_0^\prime}{\partial \Sigma}\right|_{\Sigma^*} =
\frac{(GM)^2}{(\Lambda + \Sigma^*)^3} - \frac{i}{k}\Omega =  n^* - \frac{i}{k} \Omega =
0,
\eeq
which is identical to the earlier statement of what constitutes a
step size resonance (where $n^*$ is the corresponding mean motion).
This leaves only the quadratic term,
\beq
\gamma = 
\left.\frac{\partial^2 H_0^\prime}{\partial \Sigma^2}\right|_{\Sigma^*} =
-3\frac{(GM)^2}{(\Lambda + \Sigma^*)^4} = - 3 \frac{1}{{a^*}^2}.
\eeq
The Hamiltonian can now be reduced to a standard form:
\beq
H^\prime = \frac{1}{2}\gamma\Delta\Sigma^2 + \beta \cos(k \sigma),
\eeq
where $\Delta\Sigma = \Sigma - \Sigma*$ and $\beta = - a^* S_x
(J^\prime_k(ke)/k)\cos\omega_l$.  We have also assumed that the variations
in semimajor axis are small enough that we are justified in using the
resonant semimajor axis, 
$a^*$, in $\beta$. This is just a modified pendulum Hamiltonian.
For $\beta < 0$, the argument of the cosine, $k\sigma$, oscillates
about the values $2\pi m$ (where $m$ is an integer from 0 to $k-1$).

The half-width of the resonance is normally defined by the
largest value of the momentum for which the trajectory still
librates rather than circulates.  This is
\beq
\Delta\Sigma = 2 \sqrt{\left| \frac{\beta}{\gamma} \right|} =  2
\sqrt{ S_x \frac{J^\prime_k(ke)}{3 k} (a^*)^3|\cos\omega_l| }.
\label{eq_reswidth}
\eeq
It is important to note that the resonance width depends upon $k$ but
not upon $i$.  Thus, in the vicinity of a particular step size the
closest $k:1$ resonance will be most important.  
From the definition of $\Sigma$, $\Delta L = \Delta\Sigma$.  And for
small amplitude oscillations, $\Delta a/a\sim 2\Delta L/L$
and $\Delta n/n\sim -3 \Delta L/L$.
Following the pendulum approximation further, the frequency of small
oscillations (the libration frequency) is given by
\beq
\omega^2 = \left|k^2 \beta\gamma\right| = S_x
\frac{3 k J^\prime_k(ke)}{a^*}|\cos\omega_l|.
\eeq
In the following section we use the components of the Hamiltonian
development to support
a geometric interpretation of the resonance overlap criterion.

\subsubsection{Geometric Interpretation}
\label{sec_geom}

\begin{figure}
\plotone{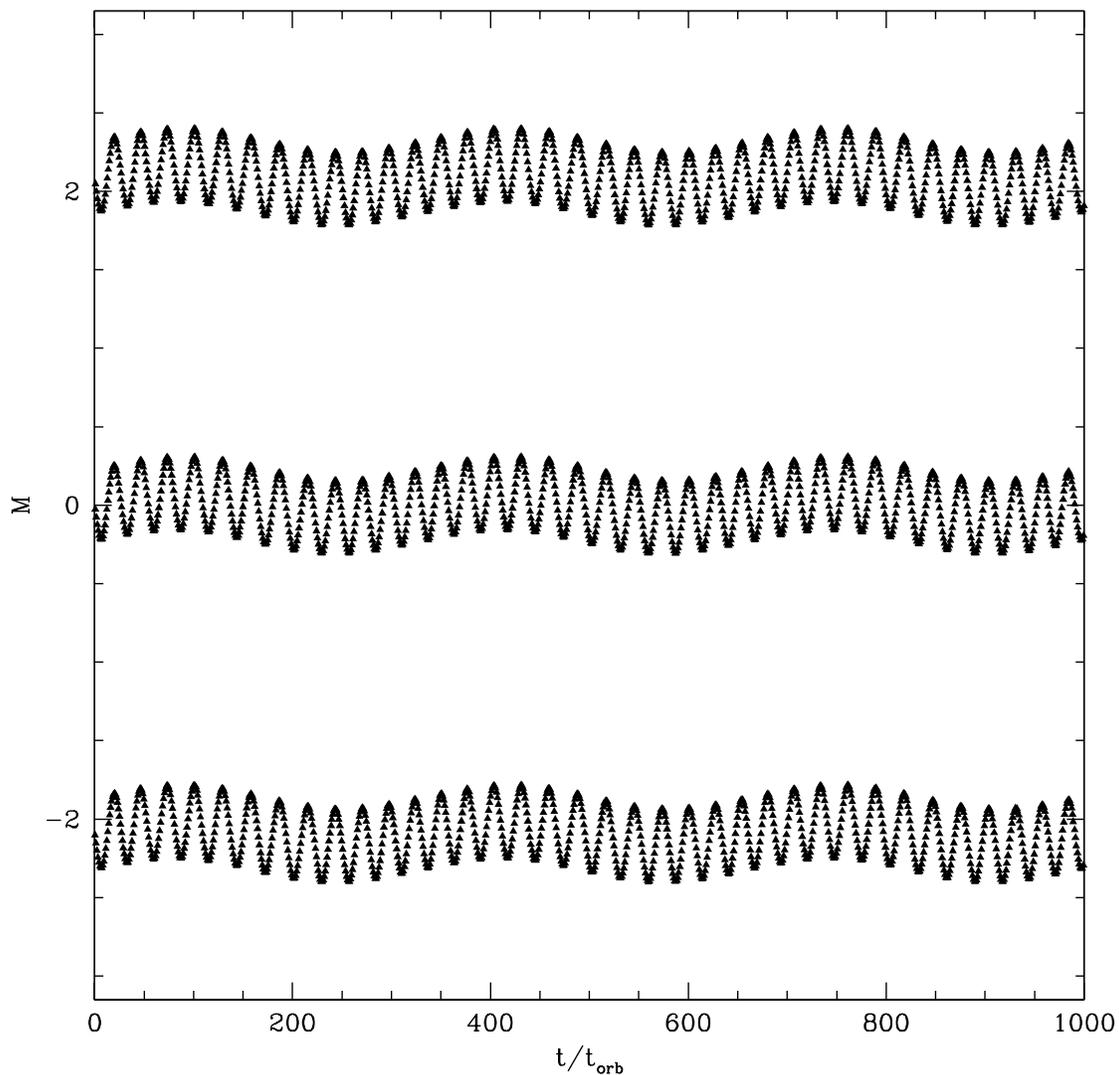}

\caption{
An example of stable libration of the mean anomaly in a 3:1 resonance. The
curves display two periods, the shorter, $\tlib$,
being the primary libration period and the longer, $t_{\rm ecc}$,
being the timescale for the eccentricity to vary between $\emin$
and $\emax$. Libration is stable only when $\tlib\lesssim t_{\rm ecc}$
and its total amplitude is $\lesssim 2\pi/k$ (for a $k:i$ resonance).
\label{fig_librate}}
\end{figure}

The libration about resonance derived above manifests itself geometrically 
in terms of an oscillation about pericenter of the orbital
mean anomaly associated with successive
integration points. As an aid in visualization, consider, for example, a 
step size equal to the mean orbital period. If the
integration is started at periapse, then each subsequent step will also begin
(and end) near periapse, aside from a small drift induced by the Stark
perturbation.
In stable libration, however, this drift away from periapse is replaced by
an {\it oscillation} centered on periapse; in this case, the particle is
stably trapped in the 1:1 resonance.
Figure~\ref{fig_librate} shows an example of libration
in the 3:1 resonance, in which each of the 3 points per orbit oscillates
around a fixed value of the mean anomaly $\cM$
(namely $\cM=-2\pi/3,\, 0,\, 2\pi/3$).
When this oscillation is stable, as it is in the figure,
the behavior shown in Figure~\ref{fig_rwalk2}
results; otherwise, the behavior is that of Figure~\ref{fig_rwalk1}.
The reason is that the stable oscillation systematically cancels out the
energy errors that would randomly accumulate in its absence, thereby
stabilizing the long-term motion (cf. the dashed curve in
Figure~\ref{fig_rwalk2}, where the
initially linear error growth turns into an oscillation on a timescale
corresponding to the libration period).
Also note that this libration cannot remain stable as $e\to 0$, 
since pericenter becomes undefined there; this qualitatively explains the
disappearance of the stable islands for $\emin\sim 0$ which was mentioned in
\S~\ref{sec_whmap}. More formally, since the width of
the $k:1$ resonance $\Delta\Sigma\propto [J'_k(ke)]^{1/2}$
(see equation~\ref{eq_reswidth}), it follows immediately that
$\Delta\Sigma\to 0$ as $e\to 0$.

With this simple picture in mind a scaling argument describing the shape of
the stable islands (Figure~\ref{fig_islands}) is easily obtained.
To do this, we first define the resonant
timestep corresponding to the $k:1$ resonance, $\dtres=\torb/k$. For a
nearby integration with timestep $\dt\sim\dtres$, define next
a ``drift'' timescale, $\tdrift$, representing the time needed for the mean
anomaly of every $k^{\rm th}$ integration point to drift by $2\pi/k$ (due to
the small mismatch between $\Delta t$ and
$\dtres$): $1/\tdrift=|1/\dtres-1/\dt|$. In qualitative terms the
libration will go unstable when $\tdrift\sim\tlib$, where $\tlib\sim
(S/\Scrit)^{-1/2}\,\torb$ can be derived from the results of the previous
section. This can be rewritten
\beq
|\Delta t-\dtres|\sim {\Delta t^2\over\torb} \left(S\over \Scrit\right)^{1/2},
\eeq
where $|\Delta t-\dtres|\ll \dtres$ and $S\ll \Scrit$ are implied.
Since $\dt\approx\torb/k$, it follows that the width of the islands should
roughly scale like $k^{-2}$; in addition, their boundaries should approximate
a parabola in the neighborhood of $\dtres$ ($S\propto (\dt-\dtres)^2$ there).
To the extent that they can be reliably measured,
these scalings hold quantitatively in 
Figure~\ref{fig_islands} (the bottom panel is cleanest in this regard).
A surprising corollary of this is that decreasing the perturbation
at fixed step size can {\it destabilize} the calculation by taking it outside
the local stable island.

We can also use the results of the previous section to estimate $\Smax(k)$,
the maximum height of the $k:1$ resonance island.
In this case we use the relations
$\Delta L/L\sim [(J'_k(ke)/k)(S/\Scrit)]^{1/2}$ and
$\Delta a/a\sim 2\Delta L/L$ (assuming small amplitude oscillations) and
note that the libration in $a$, the semi-major axis, cannot remain stable
when the $k:1$ resonance of this orbit overlaps the $k+1:1$ resonance of the
nearby orbit with semi-major axis $a+\Delta a$. The latter condition occurs
when $\dt_k(a)\sim \dt_{k+1}(a+\Delta a)$ (where $\dt_k(a)=\torb(a)/k$), which
implies $\Delta a/a\sim k^{-1}$ for $S\sim\Smax$. Equating the two then gives
\beq
\Smax(k)\sim \left(1\over 4kJ'_k(ke)\right)\Scrit.
\eeq
To order of magnitude we can take $J'_k(ke)\sim \left[(1-e^2)^{1/4}/\sqrt{2\pi
k}\right]\exp\left(-(1-e^2)^{3/2}k/3\right)$
(e.g., \cite{abrs68}; the formula becomes
exact for $e\sim 1$ as $k\to \infty$). Thus when $k\ll \kcrit=3 (1-e^2)^{-3/2}$
(i.e., when $\dt$ does {\it not} resolve periapse), $\Smax\propto k^{-1/2}$;
this is close to the actual scaling in the bottom panel of
Figure~\ref{fig_islands} (for which the empirical values of $\Smax$ are least
ambiguous). Note that the estimated $\Smax$ reaches a minimum for $k\sim
\kcrit$ and increases exponentially for $k\gtrsim\kcrit$. This suggests that the
mapping should be stable whenever periapse is well resolved;
all the numerical testing we have done confirms this prediction.
In Figure~\ref{fig_islands} this phenomenon does not occur since $\emax=1$ and
hence no fixed step size can resolve every periapse; in the 3-D case, where
$\emax<1$, a `wall' of stability is created enclosing the entire region
$\dt<\torb/\kcrit(\emax)$.
Even though the relative perturbation strength is often minuscule at periapse,
it therefore appears the stability of the WH mapping requires that it be
well resolved regardless.
The concomitant loss of efficiency for highly eccentric orbits
is obviously enormous. This motivates the search for more robust methods not
subject to this limitation, the topic to which we now turn.

\section{Modified Wisdom-Holman Mappings}
\label{sec_modwh}

There are at least two important advantages to integration methods whose
stability hinges only on resolution of the highest frequencies associated
with the perturbation forces, $\omega_{\rm pert}$,
instead of those intrinsic to the unperturbed
motion (which we presume the method to handle exactly), $\omega_{\rm orb}$.
The first, of course, is efficiency: if $\omega_{\rm orb}\gg \omega_{\rm
pert}$, such a method can use a much larger timestep---up to
$\omega_{\rm orb}/\omega_{\rm pert}$ times larger---than a scheme that must
explicitly resolve $\omega_{\rm orb}$. A more subtle advantage concerns the
unavoidable loss of energy accuracy due to round-off errors near periapse.
More specifically, if the motion is nearly-Keplerian with eccentricity $e$
no algorithm based on Cartesian phase space coordinates, that also samples
pericenter, can maintain better than $N+\log|1-e|$ digits of energy accuracy,
where $N$ is the number of arithmetic
digits carried. The proof follows immediately from
$E=p^2/2-GM/r=(rp^2-2GM)/(2r)$ and the fact that $rp^2\approx (1+e)GM$
near pericenter.
Note that doing selected intermediate calculations in extended precision
arithmetic will not improve on this; the mere act of rounding the output
values of $\mbf{x}$ and $\mbf{p}$ to $N$ digits is sufficient to do the
damage. In principle this problem can be circumvented by recasting the
dynamics in terms of osculating orbital elements (for example), but the
formulation is likely to be awkward (and probably inefficient)
since the equations are not being orbit-averaged.

The search for WH-like algorithms that are robust in the above sense is thus
well motivated.  In the following sections
we will limit ourselves to examining two recently proposed variants of the
basic WH scheme, both still exact for unperturbed Keplerian motion.
The utility of a more divergent line of methods, based on Stark instead of
Kepler motion, will be discussed in \S~\ref{sec_sint}.

\subsection{The Regularized Wisdom-Holman Mapping}
\label{sec_rwh}

\begin{figure}
\plotone{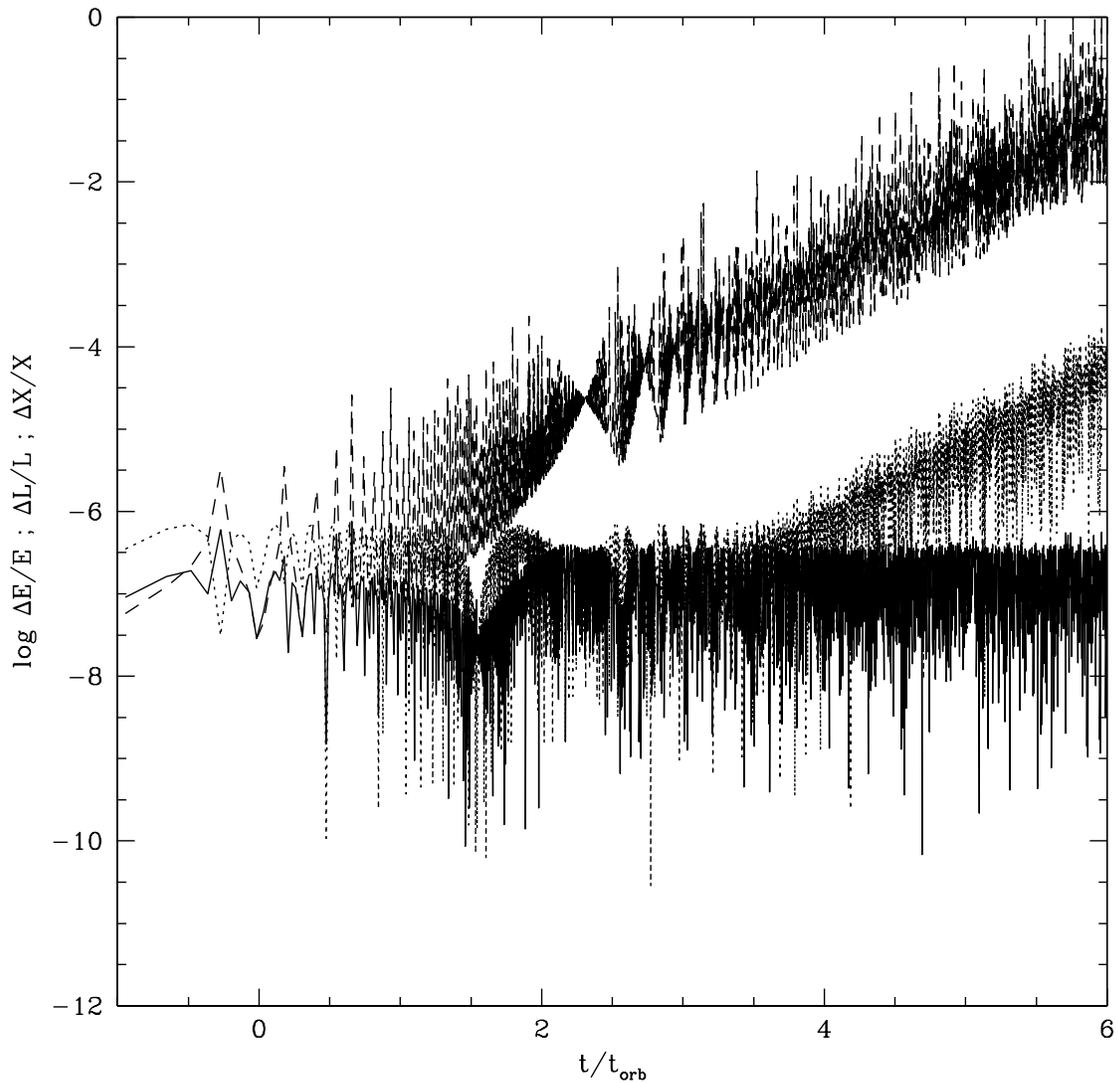}

\caption{
A typical example of the performance of the regularized
Wisdom-Holman mapping (\S~\ref{sec_rwh}) applied to the Stark problem.
The relative errors in energy (solid), angular momentum (dotted),
and position (dashed) are plotted
for an integration utilizing 100 points per orbit and the same initial
conditions and Stark vector as Figure~\ref{fig_rwalk1}. Unlike the original
mapping, the modified method is completely stable---the energy error is
bounded and the lag in orbital phase grows linearly with time.
\label{fig_rwh}}
\end{figure}

Regularization of the WH mapping through the use of an extended phase space
has been proposed by Mikkola (1997).
In this approach the integration substitutes a regularized time $s$ for the
physical time $t$, the defining relation between the two
(as in K-S regularization; e.g., \cite{stis71})
being $ds=dt/r$; further replacement of $\mbf{x}$
and $\mbf{p}$ by the corresponding K-S variables is straightforward but
entirely optional, and is not considered here. The constant steps in $s$
used by the method naturally sample pericenter more densely than constant
$t$-steps would (although it is still not quite resolved---that would require
$ds\sim dt/r^{3/2}$), offering hope of increased reliability at high
eccentricities. The substitution of $s$ for $t$ is done by extending
the phase space of the original Hamiltonian to include $t$ and $E$
(the total energy) as an additional pair of conjugate coordinates; for
details, see Mikkola (1997).

The numerical performance of this method when applied to the Stark problem
is illustrated in Figure~\ref{fig_rwh}, which plots the errors in energy,
angular momentum, and position of the method (using 100 points per orbit)
relative to the analytic solution for parameters identical to those in
Figure~\ref{fig_rwalk1}.  Accuracy and stability in this case are excellent,
as they were in every case we tried (see \S~\ref{sec_compsim} for additional
examples). Examination of the corresponding surfaces
of section confirmed that the integrated motion was regular, even for
integrations using fewer than 10 points per orbit---a remarkable feat
considering the eccentricities involved! It is in fact possible that 
the corresponding mapping Hamiltonian is integrable for arbitrary timesteps,
although we have not proven this.

This result demonstrates not only that methods which are
stable for timesteps $\dt\sim
1/\omega_{\rm pert}\gg 1/\omega_{\rm orb}$ exist, but also that they can be
simple and efficient. In fact the regularized WH map for this problem is
noticeably {\it faster}, per step, than the original mapping (cf.
Table~\ref{tab_timings} in \S~\ref{sec_compsim}); this is
because the regularized Kepler stepper does not need to solve Kepler's
Equation, and hence is much faster than the original one (this applies
only to problems of the perturbed two-body type). Because of this,
it could even be argued that the regularized method is {\it always}
superior to the
original for this class of problems, even when eccentricities are
low---although the speed increase will be noticeable only when the
cost of the Kepler step is a significant fraction of the total.
On its own, however, this method cannot handle close encounters with
perturbing objects; a promising strategy for incorporating this capability
into either the original or the regularized method is the subject of the
following section.

\subsection{Potential-Splitting Methods}
\label{sec_ps}

A well-known limitation of the WH method (and related symplectic schemes,
such as leap frog) is that the timestep cannot easily be varied without
destroying its desirable symplectic properties, such as long-term energy
conservation (e.g., \cite{gladc91}; \cite{skeg92}).
This makes the creation of adaptive
symplectic algorithms capable of handling close encounters a delicate
undertaking. Duncan, Levison, \& Lee (1997; see also Lee et al. 1997), bulding
upon the approach of Skeel \& Biesiadecki (1994), have recently developed a
symplectic, multiple-timestep generalization of the original WH method which
adds the ability to resolve close encounters without seriously compromising
its overall efficiency. We will refer to it as the `potential-splitting' (PS)
method because of the way it splits the potential of each perturber
into a series of radial zones centered around it.
We find the approach interesting not only for its
versatility in handling close encounters within a symplectic framework, but
also because the scheme is amenable to regularization. To our knowledge 
this latter possibility has not been explored elsewhere;
our subsequent discussion of the PS approach will concentrate on
determining the utility of such a merger.

In brief, the PS method works as follows. Consider for simplicity a two-body
orbit perturbed by a single point mass $m$
at a fixed position $\mbf{x}_{\rm p}$,
which generates a potential $U(\mbf{x})=-Gm/|\mbf{x}-\mbf{x}_{\rm p}|$;
the dominant central mass, $M$,
is assumed to be at the origin. This is nothing but the two fixed
point problem that will be used in \S~\ref{sec_fixed} for comparative
testing. The PS method divides the potential $U$---or more conveniently,
the perturbation force $\mbf{F}=-\nabla U$---into a series of
shells, $\mbf{F}=\sum_{j=0}^\infty \mbf{F}_j$,
each $\mbf{F}_j$ (except $\mbf{F}_0$) being non-zero over only
a finite range of $\rho=|\mbf{x}-\mbf{x}_{\rm p}|$. Introducing an ordered
sequence of radii $\rho_i$ ($i\geq -1$), where $\rho_{-1}=\infty$
and $\rho_i/\rho_{i-1}=const.<1$ ($i>0$),
one then defines $\mbf{F}_j=h_j(y)\mbf{F}$, where
$y(\mbf{x})=(\rho(\mbf{x})-\rho_i)/(\rho_{i-1}-\rho_i)$ (subject to 
$0\leq y< 1$, implying $\rho_i\leq \rho(\mbf{x})<\rho_{i-1}$) and
\beq
h_j(y)=\left\{ \begin{array}{ll}
  1-\kappa(y), & \rho_j\leq \rho<\rho_{j-1}; \\
  \kappa(y), & \rho_{j+1}\leq \rho<\rho_j; \\
  0, & {\rm otherwise},
  \end{array} \right.
\eeq
where the splitting
kernel $\kappa(y)$ is a monotonic function satisfying
$\kappa(0)=0$ and $\kappa(1)=1$. (The preceding are not the most general
definitions, but are the ones we will use.) As noted by Lee et al. (1997), 
it is also desirable for the derivatives of $\kappa$ to vanish at the
endpoints as it increases the smoothness of the transition between neighboring
$\mbf{F}_j$; they suggest $\kappa(y)=y^2(3-2y)$, which is the unique cubic
satisfying $\kappa(0)=0$, $\kappa(1)=1$, and $\kappa'(0)=\kappa'(1)=0$.

The algorithm proceeds by splitting the base timestep $\Delta t$ into
an integral number of smaller steps whenever the test particle 
enters a zone interior to the one it previously resided in---i.e., whenever it
happens to approach the perturber sufficiently more closely. At each such
subdivision, a specific kick component $\mbf{F}_j$ is applied in such a way as
to allow further subdivisions if necessary yet keep the entire process
symplectic; eventually all higher numbered $\mbf{F}_j$ vanish, the
recursion is terminated, and the
intervening Kepler drift about $M$ is performed.
The net effect is to take many small steps during close encounters; far from
the perturber, the method reduces to the standard WH mapping (albeit in
kick-drift-kick form instead of drift-kick-drift form; cf.
equation~[\ref{eq_sint_step}]).  For further details, consult the references.

We now propose several enhancements to the basic PS method which we have found
can significantly improve its stability. The first is a modified kernel
function, $\kappa(y)$. The motivation is simply to produce a fully analytic
decomposition of $\mbf{F}$, meaning one for which {\it all} derivatives of
$\kappa(y)$ vanish at the endpoints, instead of merely the first; this is
conceptually similar to the creation of a ``$C^\infty$ bump function.''
Many forms are possible; the one we have found most useful is
\beq
\kappa(y)={1\over 2}\left\{1+\tanh\left[2y-1\over y(1-y)\right]\right\}.
\eeq
Although more costly to evaluate than the original polynomial (the new
code was about 30\% slower), the resulting method was stable for larger $\dt$
and longer periods of time than the original in every test tried (including
the two fixed center problem, \S~\ref{sec_fixed}). On the other hand,
since in many cases
the polynomial kernel performed nearly as well, we do not claim our
modified kernel is uniformly superior; we have, however, found it a useful
alternative in situations where the original seems to behave poorly.

The second enhancement involves force-center switching. Whereas the basic PS
method takes Kepler drifts about $M$ regardless of the test particle's
proximity to the perturber $m$, it is clearly advantageous to execute drifts
about $m$ during very close encounters. This capability can be neatly
incorporated into the PS scheme as follows. Note first that the potential $U$
is not split at all when $\rho>\rho_0$; this is the regime in which
$\mbf{F}=\mbf{F}_0$, where no subdividing is done and
the method is equivalent to the usual WH approach.
Now define a value $\jmax$ such that $\mbf{F}=\mbf{F}_{\jmax}$ whenever
$\rho<\rho_{\jmax}$; this implies that all $\mbf{F}_j$ are identically zero
for $j>\jmax$. Thus when $\rho<\rho_{\jmax}$ the full Hamiltonian is again
available and can be separated into a piece representing an $m$-drift and a
piece representing the ``perturbation'' $M$ with potential $V=-GM/|\mbf{x}|$.
Next, modify the rules for subdivision so that when
$\rho<\rho_{\jmax}$, kick($V$/2)-drift($m$)-kick($V$/2) is done
instead of kick($U_{\jmax}$/2)-drift($M$)-kick($U_{\jmax}$/2); since
$U_{\jmax}=U$ here both forms are equally valid. The resulting method now
uses drifts around $m$ (without further subdivision of the timestep)
whenever the encounter is
close enough. Although any $\jmax>0$ may be used, it should obviously be
chosen so that $m$ dominates the dynamics for $\rho<\rho_{\jmax}$. It may
appear that switching splittings in this fashion breaks the symplecticity,
and this is entirely possible; if true, however, we have not found it to
be a problem. In tests with the two fixed center problem, for example, energy
conservation remained stable even after thousands of switchings. This may be
due partly to the fact that the local timestep is very small when the
switching takes place, minimizing any systematic errors it may commit.
The technique appears a promising one in any event.

The final enhancement we considered was the incorporation of regularization
into the PS framework. This is in fact quite straightforward to accomplish;
one simply splits the regularized Hamiltonian instead of the original one---the
$\mbf{F}_j$ becoming regularized force components, and so on. The usefulness
of doing this is also clear: whereas the basic PS method shares the
instability of the WH mapping it is based on when orbits are eccentric, the
regularized PS method remains robust here. We have verified this directly in
the case of the Stark problem; in particular, the instability in the
unmodified PS method was found to
persist even when splitting of the Stark potential at small radii was
included---not to mention the fact that doing this made the method extremely
inefficient! As we will clearly demonstrate in \S~\ref{sec_compsim}, the
regularized PS algorithm appears completely robust in this regard.

\section{Stark-based Integration Schemes}
\label{sec_sint}

A more radical strategy for creating integrators reliable in the way
outlined in \S~\ref{sec_modwh} is not to make minor transformations to
the original Hamiltonian splitting,
but instead to rethink the Kepler splitting entirely.
In this case the primary motivation for altering the splitting 
is not to produce a faster method, as was true for the WH mapping
(compared with leap frog, say),
but rather one that is more stable. The challenge is that each piece of the
new Hamiltonian splitting should be integrable and efficiently soluble if the
mapping is to be practical. Although integrable problems in general are a
precious commodity, the preceding analysis provides
two obvious candidates: the Stark problem and the two fixed point problem.
(The two are in fact closely related, since the latter reduces to the Stark
problem in the limit $|\mbf{x}_{\rm p}|\propto m^{1/2}\to \infty$, where $m$
and $\mbf{x}_{\rm p}$ are the mass and position of one of the fixed points.)
In this paper, however, we will only consider mappings based on a Stark
splitting of the Hamiltonian, arguably the simpler of the two (it having one
less free parameter).
For comments on the use of the two fixed point problem as the basis of a
symplectic mapping, refer to \S~\ref{sec_discuss}.

\subsection{The Time-Reversible Stark Method}
\label{sec_trs}

The first Stark-based method considered was derived from the original WH
mapping by simply replacing the Kepler stepper in that method with a Stark
stepper, where the value of the Stark vector for a given step was 
self-consistently taken equal
to the value of the perturbation force at midstep; consequentially,
the intervening `kick' vanishes.
(Hence computation of the Stark vector requires iteration, at least
in principle.) An integration therefore consists solely
of a sequence of Stark steps, the Stark vector for any particular step
representing a local best-fit to the perturbation. 
Note that it would be misleading to think of this method as symplectic. The
basic problem is that in each step the Stark vector depends on the current
position of the test particle, yet this coordinate dependence in the
mapping Hamiltonian is not accounted
for by the Stark step itself. Thus although it is perfectly reasonable to
think of any one step as being symplectic---since a strictly constant
Stark vector which cancels the corresponding kick 
{\it for that step} is easily manufactured---the unconstrained manner in which
the vector changes destroys the self-consistency required
for the {\it sequence} of successive steps to exhibit symplectic behavior.
The situation is closely analogous to attempting to vary the step size in,
for example, leap frog or the WH method: although any one step is obviously
symplectic, successive steps are not part of the {\it same} Hamiltonian flow
and hence the integration does not display coherent symplectic properties such
as long-term energy conservation. By construction, however, the
method is explicitly time-symmetric and hence might still exhibit good energy
behavior; we will refer to it as the time-reversible Stark method. Also
note that for eccentric orbits the perturbations near pericenter will
generally be nearly stationary, so that the Stark approximation should be an
excellent one here---giving hope that the scheme will maintain stability in
such cases. Disappointingly, this hope was quickly dashed by the results
of our numerical tests; for details, see \S~\ref{sec_compsim}.

\subsection{Regularized Stark Mappings}
\label{sec_rss}

The regularization and extended phase space techniques employed by Mikkola
(1997) can also be harnessed to create a regularized, fully symplectic
Stark-based mapping. Consider in particular a time-dependent, perturbed two-body
Hamiltonian of the form $H=\mbf{p}^2/2-1/r+U(\mbf{x}, t)$ (where $r=|\mbf{x}|$),
and introduce a fictitious 
Stark potential $-\mbf{S}(t)\cdot\mbf{x}$ that reproduces $U(\mbf{x}, t)$ as
closely as possible; the critical restriction
is that $\mbf{S}$ can depend only on $t$,
{\it not} on $\mbf{x}$. For example, if $U(\mbf{x},
t)=-Gm/|\mbf{x}-\mbf{x}_{\rm p}(t)|$, then $\mbf{S}(t)=Gm\mbf{x}_{\rm p}(t)/|
\mbf{x}_{\rm p}(t)|^3$ would be optimal whenever $|\mbf{x}|\ll |\mbf{x}_{\rm
p}(t)|$ (and otherwise the perturbation is not Stark-like at all, so using a
Stark splitting would be pointless to begin with!) Now rewrite $H$ as
\beq
H(t)={\mbf{p}^2\over 2}-{1\over r}-\mbf{S}(t)\cdot\mbf{x}+\delta U(\mbf{x}, t),
\eeq
where $\delta U(\mbf{x}, t)=U(\mbf{x}, t)+\mbf{S}(t)\cdot\mbf{x}$. Letting
$\cE=-H(0)$ be the total binding energy and extending phase space using
$ds=dt/r$ (cf. \S~\ref{sec_rwh}), the extended, regularized Hamiltonian is
\beq
{\tilde H}=r\left[{\mbf{p}^2\over 2}-{1\over r}-\mbf{S}(t)\cdot\mbf{x}\right]+
r\left[\cE+\delta U(\mbf{x}, t)\right].
\eeq
The grouping is intended to show how ${\tilde H}={\tilde H}_0+{\tilde H}_1$
is to be split. The first
piece, ${\tilde H}_0$,
is clearly just the (regularized) Hamiltonian for the time-dependent
Stark problem; however, since ${\tilde H}_0$
is independent of $\cE$ Hamilton's equations
imply that the coordinate $t$ is a {\it constant} here (it is advanced only
by the
second piece of the Hamiltonian, ${\tilde H}_1$). Therefore, $\mbf{S}(t)$
is a constant vector for the duration of the ${\tilde H}_0$ step, and we have
succeeded in creating an obviously symplectic algorithm based on
perturbed Stark motion instead of perturbed Kepler motion. In addition, the
(optional) regularization used can be expected to
provide the same protection against
instability as occurred in the regularized WH method.

Although we came upon it independently, the above method
({\it without} regularization)
is very similar to the one created by
Newman et al. (1997; see also \cite{gra97}). Their motivation, by contrast,
was in exploring its use in treating close encounters;
their approach (as we understand it) represents a modification of the method of
Levison \& Duncan (1994) in which
the Stark splitting is employed only during close encounters, 
just {\it after} switching force centers to the nearby perturber.
Irrespective of motivation, it must be kept
in mind that there is a significant price to pay for using a Stark splitting,
since Stark steps (in our experience) are roughly 30 times
slower than the Kepler steps they are replacing. Unless the force calculation
strongly dominates the execution time, this implies that an increase in the
step size by a factor of 30 must be permissible for the change to pay
off. This is clearly a severe constraint! On the other hand, if the method is
stable for $\dt\sim 1/\omega_{\rm pert}$, and $\omega_{\rm orb}/\omega_{\rm
pert}\gtrsim 30$ (cf. \S~\ref{sec_modwh}), then such huge improvements would be
plausible.
%(but as noted above, the method does {\it not} appear to be stable
%in this way).
When the original Kepler mapping is also stable, however, it is
quite unlikely gains of that magnitude would be possible---particularly since
the use of symplectic correctors (\cite{wisht96}), which we have not taken
advantage of, can often substantially
boost the accuracy of the original mapping with no loss in overall efficiency.

\begin{deluxetable}{ccccc}
\tablecaption{Relative Efficiency of Selected Integration Simulations
\label{tab_timings}}

\tablehead{
\colhead{Integrator} &
\colhead{Fig.~\ref{fig_fixcmp1}} &
\colhead{Fig.~\ref{fig_fixcmp2}} &
\colhead{Fig.~\ref{fig_fixcmp3}} &
\colhead{Fig.~\ref{fig_bhcmp}}
}

\startdata
  WH&  1.00& 1.00& 1.00& 1.00\nl
  RWH& 0.75& 0.73& 0.74& 1.40\nl
  PS&  1.20& 14.5& 6.40& 17.5\nl
  MPS& 1.25& 20.3& 8.50& 25.5\nl
  TRS& 12.8& 15.0& 20.0& 12.8\nl
  RSS& 13.7& 11.9& 13.8& 11.1\nl
\enddata

\end{deluxetable}

\section{Comparative Simulations}
\label{sec_compsim}

\subsection{The Two Fixed Point Problem}
\label{sec_fixed}

Having described four alternatives to the original WH scheme---
the regularized WH mapping, the PS method (with and without our purported
enhancements), and two types of Stark-based schemes---we now wish to provide
some practical insight into their relative performance. In this section the
two fixed point problem, whose dynamics is understood in complete detail, is
used as a test problem; in \S~\ref{sec_gdtest} a more generic test problem
from the area of galactic dynamics is used to estimate performance under more
``typical'' conditions.

\subsubsection{Orbital Motion}
\label{sec_forb}

\begin{figure}
\plotone{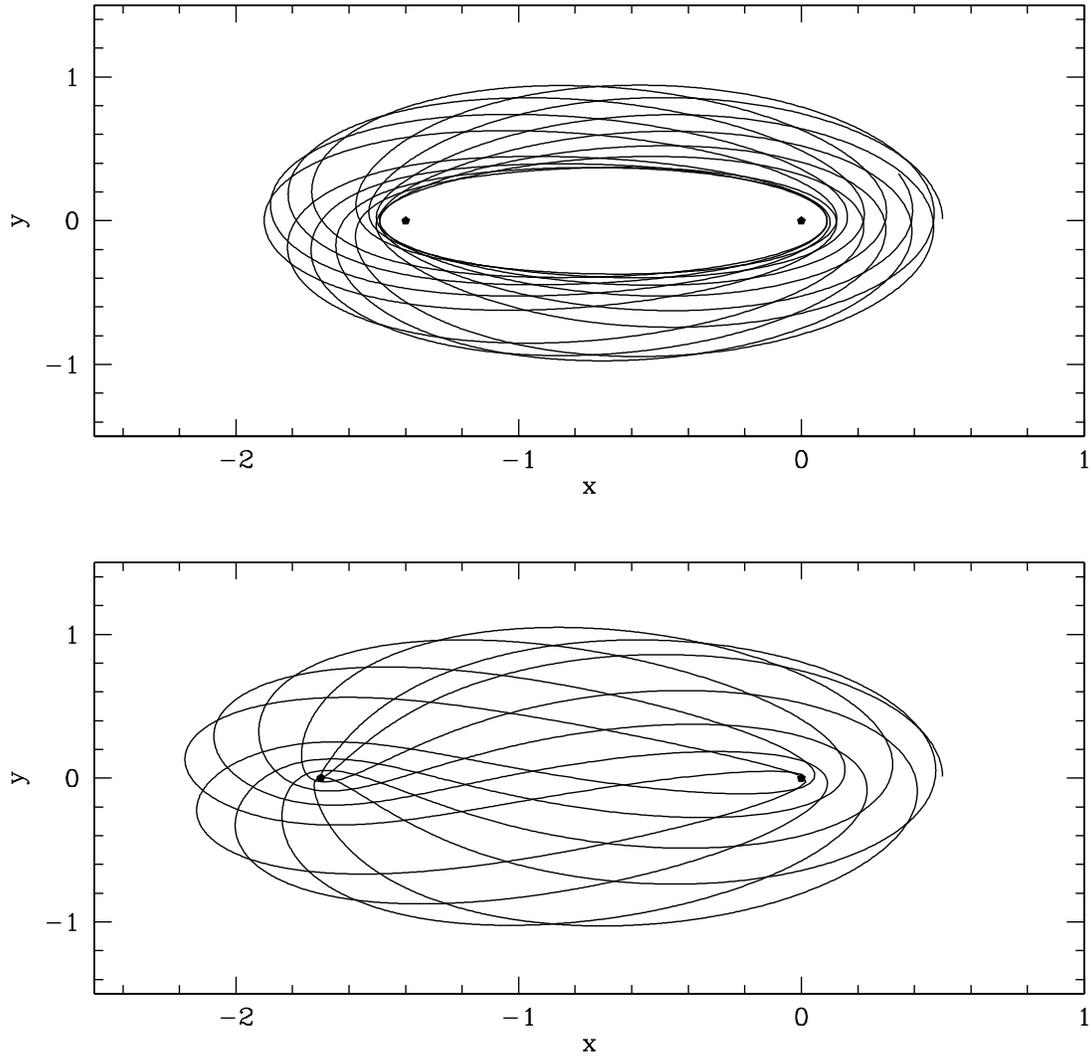}

\caption{
Two examples of bounded motion in the two fixed point problem
(\S~\ref{sec_fixed}). In the top panel, the motion is confined 
to lie between two ellipses (with foci at the indicated
positions of the fixed masses)
and hence maintains a finite distance from both masses. In the bottom panel,
the inner ellipse no longer exists and arbitrarily close (and radial)
encounters with both masses are possible.
\label{fig_fixed}}
\end{figure}

The two fixed point problem (e.g., \cite{par65}) represents the motion of a
test particle in the field of two gravitating
point masses, $m_1$ and $m_2$, held at fixed positions, $\mbf{x}_1$ and
$\mbf{x}_2$; the Hamiltonian per unit test mass is
\beq
H={\mbf{p}^2\over 2}-{Gm_1\over |\mbf{x}-\mbf{x}_1|}-
	       {Gm_2\over |\mbf{x}-\mbf{x}_2|}.
\eeq
Like the Stark problem, it is fully integrable and possesses
three constants of motion; as noted earlier, it in fact reduces to the Stark
problem in the limit (for example) $|\mbf{x}_2|\propto m_2^{1/2}\to \infty$.
The problem is separable in confocal coordinates and can be solved
analytically in terms of elliptic functions and integrals.
The three constants of motion are the energy $E$, the angular momentum
component along the direction $\mbf{x}_2-\mbf{x}_1$, and a separation constant
$\alpha$ (an explicit formula for which can be found in Lessnick (1996)).

Although the problem is rather artificial (we can think of no good physical
analogies), it does allow close encounters to be introduced in a controlled
manner. In this paper we will consider only two-dimensional motion
($p_z=z=z_1=z_2=0$, say) with the further specializations $\mbf{x}_1=0$,
$\mbf{x}_2=(x_{\rm p}, 0)$, and $Gm_1=1\gg Gm_2$, so that the motion is
nearly-Keplerian except very 
near $m_2$. There are three classes of bound motion (unbound orbits will not
be considered).  First, the particle can be tightly bound to either $m_1$ or
$m_2$, never closely approaching the other mass; this is the limit
in which the motion is (quantitatively) similar to that in the Stark problem.
In the second type of motion, the test body is confined to the annulus between
two confocal ellipses (with foci at the positions of $m_1$ and $m_2$) and
hence maintains a finite distance from both masses at all times. In the third
type of motion, the inner bounding ellipse disappears and the particle
eventually approaches each mass arbitrarily closely and with encounter
eccentricities arbitrarily close to unity. Examples of the latter two motions
are shown in Figure~\ref{fig_fixed}.

We shall focus attention on
the integration of initial conditions lying near the critical line between
motion of the second and third types; on the critical curve, the inner ellipse
degenerates to the line connecting $m_1$ and $m_2$ and the limiting motion
consists of an ever-tightening spiral which converges on this line.
Convenient, precisely critical initial conditions are $\mbf{x}_0=(1, 0)$,
$\mbf{p}_0=(0, 1)$, and $x_{\rm p}=-1$ ($m_2$ remains arbitrary);
slight variations of $x_{\rm p}$ from this critical value then allow exquisite
control over the severity of close encounters.

\subsubsection{Numerical Integrator Performance}
\label{sec_fint}

\begin{figure}
\plotone{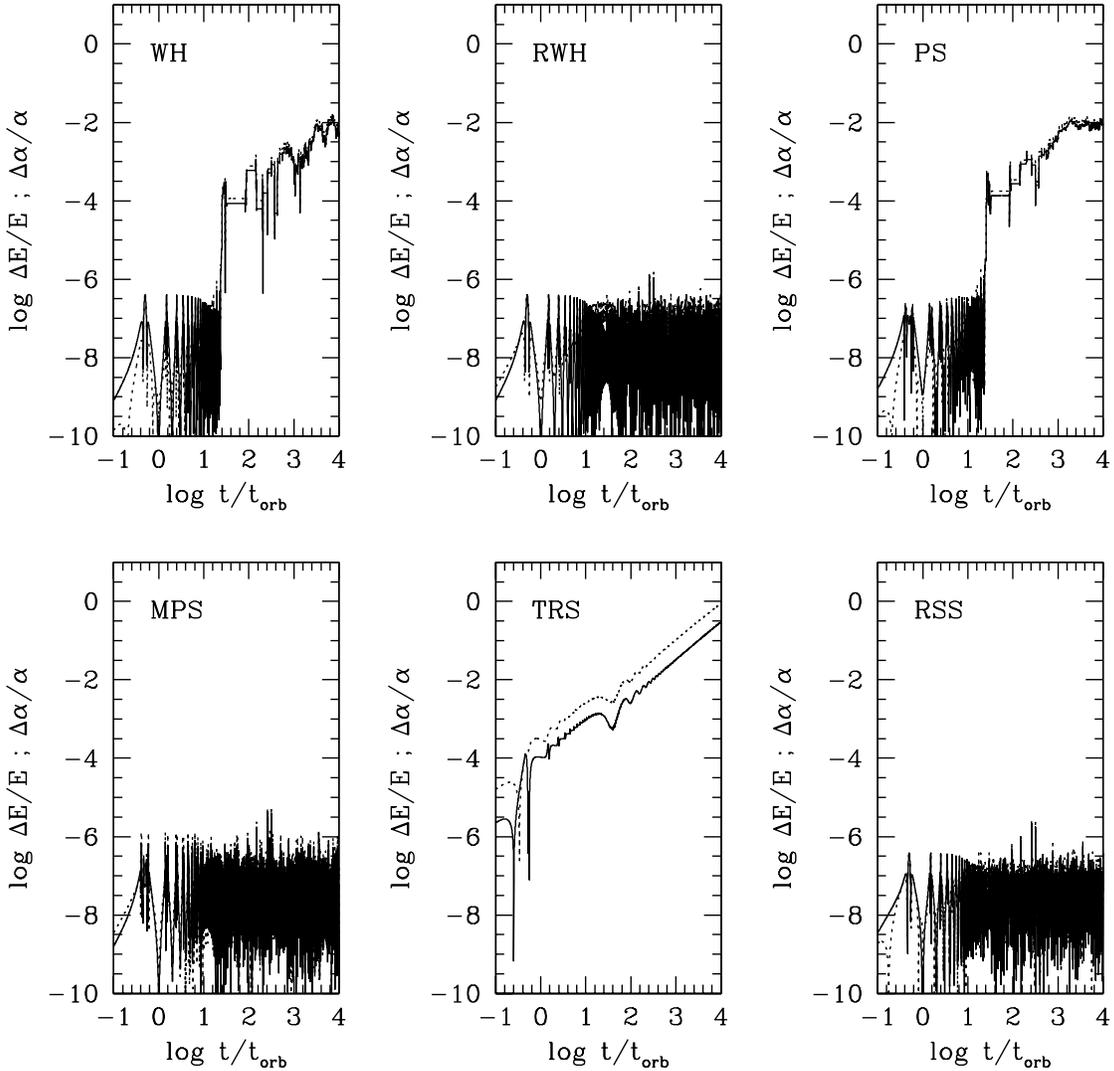}

\caption{
Comparative integrator performance for the two fixed point problem with
$x_{\rm p}=-1.5$ and $Gm_2=0.01$ (see \S~\ref{sec_fint}). In this case there
are no close encounters with the mass $m_2$ and the orbital
motion is similar to that in the Stark problem (cf. Figure~\ref{fig_sorb}).
Each panel displays the relative errors in energy $E$ (solid curve) and
separation constant $\alpha$ (dotted curve) for a specific integration scheme.
All regularized methods (RWH, MPS, RSS) are stable and well-behaved; all
others go unstable when the orbit becomes nearly radial. The TRS scheme
also exhibits linear error growth.
\label{fig_fixcmp1}}
\end{figure}

\begin{figure}
\plotone{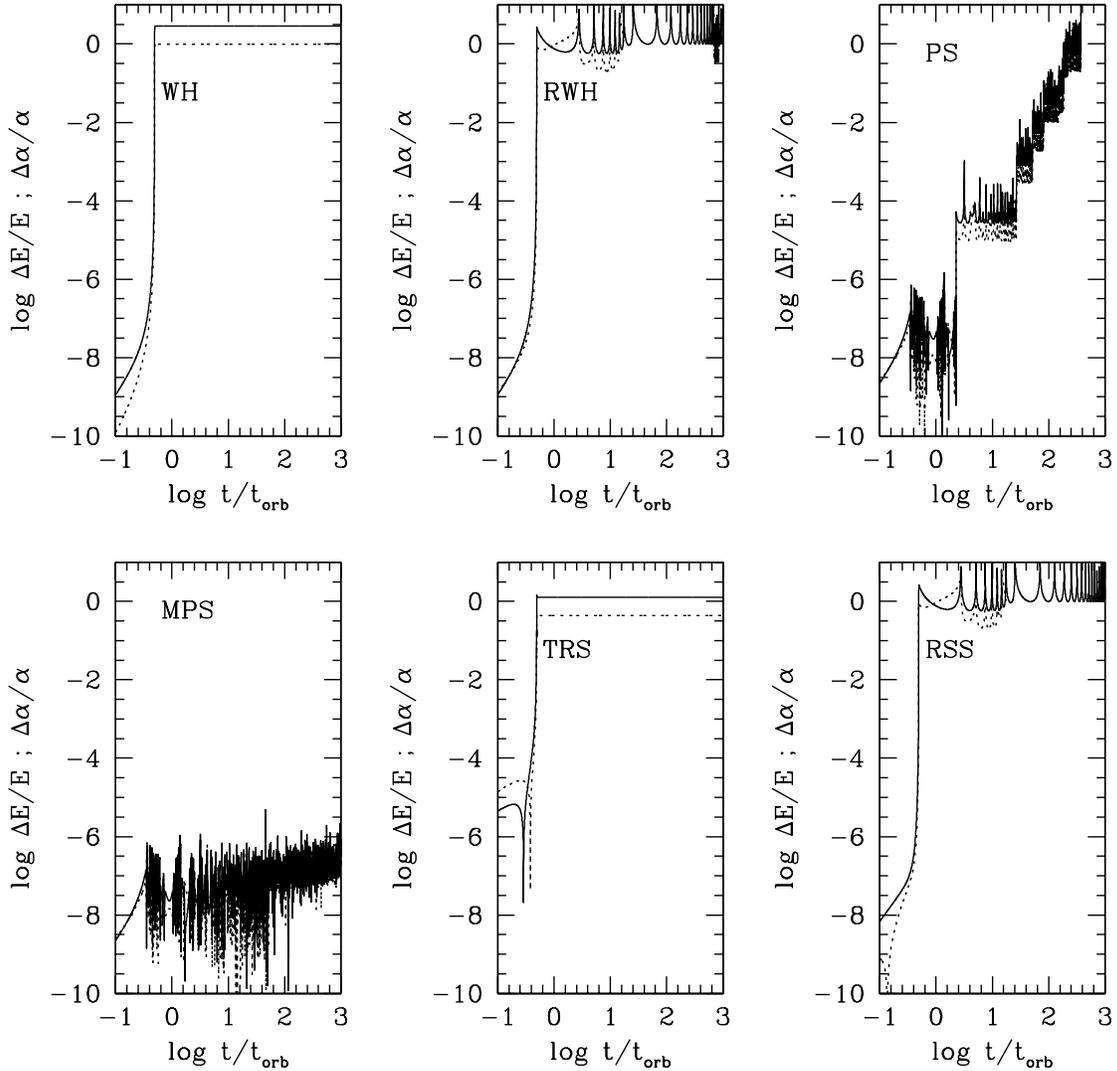}

\caption{
Similar to
Figure~\ref{fig_fixcmp1} but for $x_{\rm p}=-1.02$. Here the orbital motion
is like that in the lower panel of Figure~\ref{fig_fixed} and the test particle
undergoes arbitrarily close (and radial) encounters with both masses.
As expected, all the single-timestep integrators (WH, RWH, TRS, RSS) quickly
falter since they cannot handle close encounters. Although
initially stable, the PS integrator (here {\it including} regularization)
is occasionally overwhelmed by the 
extremely close encounters present in the problem. Only the
MPS algorithm, incorporating both regularization and
force-center switching, performs satisfactorily.
\label{fig_fixcmp2}}
\end{figure}

\begin{figure}
\plotone{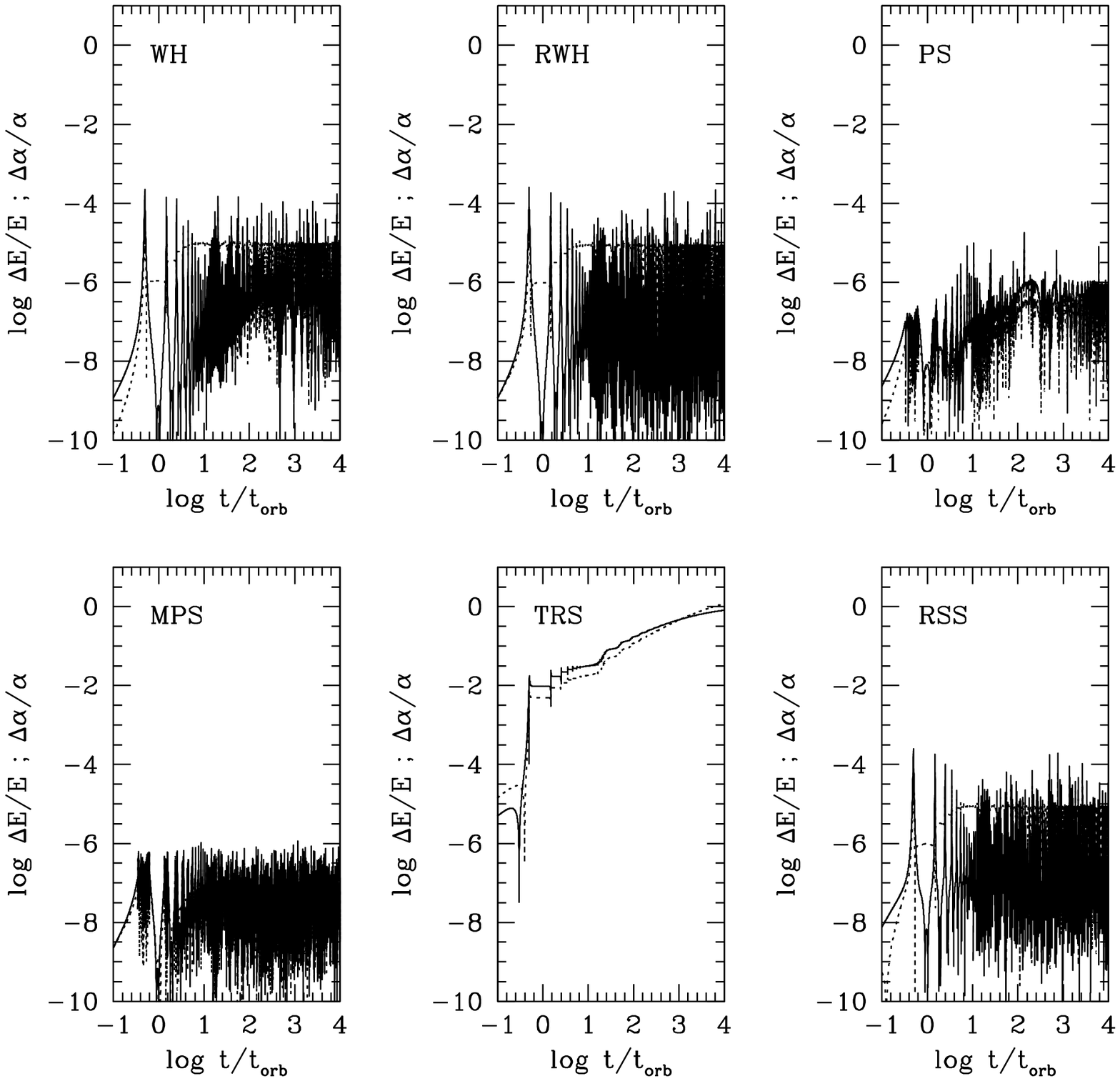}

\caption{
Similar to
Figure~\ref{fig_fixcmp1} but for $x_{\rm p}=-0.95$. The orbital motion now
corresponds to the top panel of Figure~\ref{fig_fixed}: both masses are
approached, but encounter distance and eccentricity both have finite bounds
($b\gtrsim 0.05$ and $e\lesssim 0.95$, respectively).
The chosen timestep, $\dt=10^{-3}\torb$, was sufficient to resolve all close
encounters and hence all integrators (except the TRS method) perform well.
As in Figure~\ref{fig_fixcmp1}, the TRS scheme displays linear error growth.
\label{fig_fixcmp3}}
\end{figure}

Numerical results are shown in Figures~\ref{fig_fixcmp1}-\ref{fig_fixcmp3};
the relative execution times for each simulation are given in
Table~\ref{tab_timings}.
Each figure contains six panels, one for each integration method tested:
the original Wisdom-Holman method (labeled `WH'), the regularized WH mapping
(`RWH', \S~\ref{sec_rwh}),
the original potential-splitting method (`PS'), our modified PS
algorithm (`MPS', \S~\ref{sec_ps}), the time-reversible Stark scheme (`TRS',
\S~\ref{sec_trs}), and the regularized, symplectic Stark method (`RSS',
\S~\ref{sec_rss}). The Stark stepper for the TRS and RSS integrators
implemented the analytic solution in the case of bound motion;
for unbound motion,
a highly accurate (machine level truncation error) Bulirsch-Stoer routine was
used to numerically integrate the K-S regularized equations of motion.
Each panel plots the relative errors in the energy $E$
(solid curve) and separation constant $\alpha$ (dotted line) over the course
of an integration lasting $10^3$ or $10^4$ orbital periods, where $\torb$
is the period of the unperturbed Kepler orbit. (Since in most cases the errors
in $E$ and $\alpha$ are very similar, the two curves are often difficult to
distinguish.) All integrations used $\Delta t=10^{-3}\torb$, $Gm_2=0.01$, and 
the aforementioned near-critical initial
conditions; only the value of $x_{\rm p}$ was changed between figures.

Figure~\ref{fig_fixcmp1} plots the results for $x_{\rm p}=-1.5$. In this case
there are never close encounters with $m_2$ and the motion is similar to that
in the Stark problem; in particular, the orbit periodically becomes highly
eccentric. The qualitative results in this case are quite simple---all
regularized schemes (RWH, MPS, RSS) behaved well and showed no signs of
instability, whereas all unregularized ones proved to be unstable at high
eccentricities. In addition, although the TRS scheme performs well initially
(which is not surprising, since the motion is Stark-like) its errors grow
secularly and by the end of the simulation are of order unity; this is
disappointing considering that the method is explicitly time-symmetric, and
further supports the `non-symplectic' label previously placed on it (see
\S~\ref{sec_trs}).

Figure~\ref{fig_fixcmp2} shows the results for $x_{\rm p}=-1.02$. In this case
there are frequent close encounters with both masses and there is
no limit to how close and radial they can be (cf. Figure~\ref{fig_fixed},
lower panel). 
As expected, all the single-timestep integrators (WH, RWH, TRS, RSS) quickly
falter, unable to cope with the close encounters.  Although
initially stable, the PS integrator is occasionally overwhelmed by the 
extremely close encounters present in the problem; further investigation
revealed that the discrete jumps occurred during encounters with impact
parameters $b\lesssim 10^{-3}$.
(In this instance the PS integrator was regularized since otherwise
it would have been unstable even in the absence of close encounters.)
Only the MPS algorithm, which also included force-center switching,
performs satisfactorily. As evidenced by the formation of a secular trend late
in the test, however, even this algorithm has its limits. In particular we
have not found a practical way to regularize {\it about the perturber} while
the force-center switch is active, and hence the scheme is unstable
whenever the encounter eccentricity exceeds some limit. The growing
error may also be partly due to non-symplectic behavior (or `ringing')
introduced during the switching process, but as both effects occur only during
encounters it is difficult to distinguish between the two (at least in these
simulations).

The results for $x_{\rm p}=-0.95$ are displayed in Figure~\ref{fig_fixcmp3}.
Here the qualitative orbital motion corresponds to the top panel of
Figure~\ref{fig_fixed}; although both masses are approached, all encounters
have impact parameters $b\gtrsim 0.05$ and local eccentricities
$e\lesssim 0.95$. This implies that the chosen timestep ($\dt=10^{-3}\torb$)
is just adequate to clearly resolve the encounters with both $m_1$ and $m_2$.
In this case, therefore, nearly every integrator performs quite well.
The exception is the TRS method, which is again plagued by a strong
linear error growth.
The PS and MPS routines do particularly well because of the effectively
smaller timestep used near $m_2$ (the approaches to $m_2$ are close enough
that the algorithms subdivide the timestep several times each encounter),
but the relative cost in execution time (see Table~\ref{tab_timings})
is commensurate---the WH and RWH schemes would have
produced comparable or superior results given the same amount of CPU time.

The timing results listed in Table~\ref{tab_timings} are straightforward to
interpret. The RWH method is the most efficient due to the speed of the
regularized Kepler stepper, the force calculation being essentially trivial
here. The PS and MPS methods show little overhead cost in the absence of close
encounters, but slow down substantially when close approaches occur frequently.
The rather obvious conclusion is that multiple-timestep
integrators (and other `encounter codes')
should only be used when the detailed encounter dynamics are
important---if a moderately softened perturbing potential is acceptable,
for example, integration using a constant timestep scheme would be both stable
and more efficient. The Stark-based schemes are at least an order of magnitude 
slower than the corresponding Kepler mappings because of the high cost of
taking Stark steps instead of Kepler steps, yet in this case
were no more accurate. We conclude that such methods are uncompetitive unless
the perturbation potential is {\it extremely} Stark-like; in particular,
for point mass perturbers---which only appear Stark-like at large distances
and to lowest order---we consider them to be of marginal interest.

\subsection{Galactic Dynamics Test Problem}
\label{sec_gdtest}

\begin{figure}
\plotone{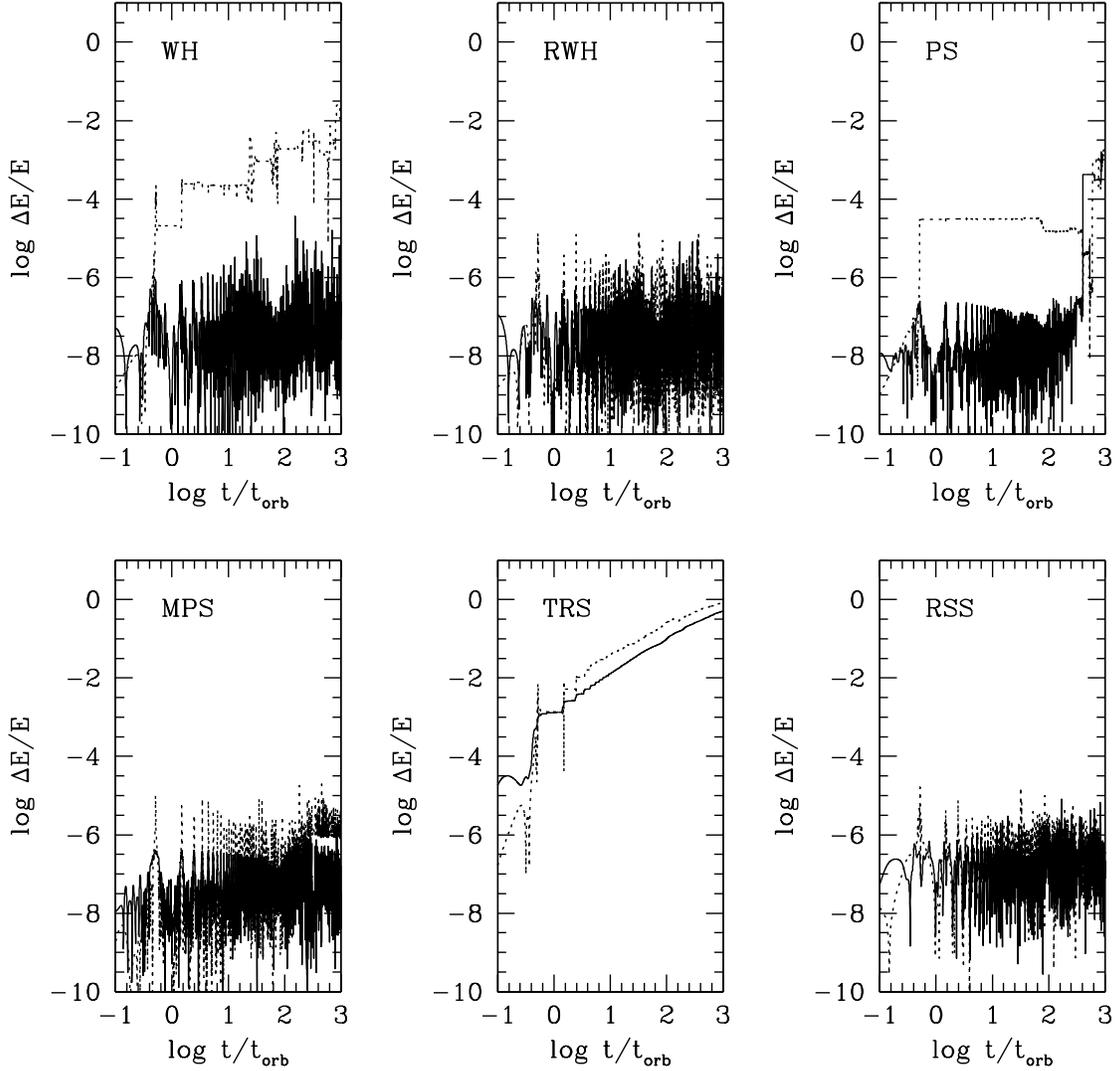}

\caption{
Results for the galactic dynamics test problem (\S~\ref{sec_gdtest}; cf.
Figures~\ref{fig_fixcmp1}-\ref{fig_fixcmp3}). The solid curve shows the energy
error for an orbit with low initial eccentricity ($e_0=0.5$); the dotted
line corresponds to $e_0=0.99$ (the rest of the configuration remaining
unchanged). The performance of each integrator is in line with expectations;
in particular, only the regularized methods remain stable when $e\sim 1$, and
the TRS method continues to be dominated by a secular trend.
\label{fig_bhcmp}}
\end{figure}

The results of a more generic test problem are shown in
Figure~\ref{fig_bhcmp} (relative timings can again be found in
Table~\ref{tab_timings}). The problem consisted of integrating test particle
motion in the perturbing 
field of 100 fixed points of mass $10^{-3}M$ (where $M$ is
the mass of the central object, held fixed at the origin).
The positions of the masses were drawn randomly from a
spherically symmetric distribution with a radial density profile
$\propto r^{-2}$; this is similar to the mass distribution seen in several
galactic nuclei believed to contain massive black holes, and closely
resembles the configurations used by Rauch \& Ingalls (1997).
(We do not allow the perturbers to orbit $M$ because energy would not
be conserved in this case.) In addition,
for the WH, RWH, TRS, and RSS integrators (i.e., the constant timestep
methods) the perturbing potentials were slightly softened to limit the
severity of close encounters and allow a more realistic comparison with the
multiple-timestep (PS and MPS) routines to be made.
The step size in all cases was $\dt=10^{-3}\torb$.
Finally, we note that the MPS integrator used in these simulations
did not include force-center
switching, which would have required substantial (though conceptually
straightforward) modifications to the original code due to the multitude of
perturbers involved.

As before each panel in the figure plots the relative energy error of 
a particular integrator. The solid line represents the results for an orbit
with low initial eccentricity ($e_0=0.5$),
and the dotted line for one with high eccentricity
($e_0=0.99$); the initial conditions were otherwise identical and were
the same for every integrator. The results are completely in line with
previous ones and show nothing unexpected. Only the regularized methods are
stable at high eccentricities, the PS and MPS algorithms appear highly
successful at resolving the many ({\it unsoftened}) close encounters that
occur, and once more the error in the TRS scheme is dominated by a secular
drift.

\section{Discussion}
\label{sec_discuss}

We have shown that the WH mapping is generically unstable when applied to
eccentric, nearly-Keplerian orbits whenever the step size is not small
enough to resolve periapse. This `radial orbit instability' is fully
explainable in terms of the overlap of step size resonances and has a simple
geometric manifestation in the case of the Stark problem. Our investigation
indicates that the islands of stability found in the latter problem do not
exist in the more general cases we have examined; the instability therefore
appears to be unavoidable in typical situations, unless one employs the
brute-force approach of decreasing the timestep by the requisite amount.
However, besides the fact that this is an extremely inefficient solution---it 
reduces the mapping to a very costly direct integration scheme---we have shown
that an elegant solution to the problem is already available: the
regularization approach of Mikkola (1997). In every case examined, not only
was the regularized WH mapping immune to the radial orbit instability,
in many cases it was also more efficient.
We enthusiastically recommend its use whenever close encounters with
perturbers are not of concern.

We remind the reader that our investigation has not cast into doubt
all previous studies that have used the WH integrator and its
variants.  In nearly every case care has been taken to use a small enough step
size such that perihelion passage would be adequately sampled.  We
note only one area where particular caution should be exercised.  One
of the features of the long-term dynamics in mean motion resonances
and secular resonance is that very high eccentricities can be
developed.  These eccentricities are often large enough that physical
collision with the sun is a common outcome in studies of meteorite
delivery from the main asteroid belt and the long-term dynamics of
ecliptic comets (\cite{glaet97}; \cite{morm95}; \cite{levd97}).
In those cases it is unlikely that the step size used
was small enough to resolve the perihelion passage.    Although these
researchers checked their results for step size dependence and
reported no numerical artifacts, we suggest that further examination
of those cases would be prudent.

We have demonstrated that the potential-splitting method of Lee et al. (1997)
can be regularized to produce an algorithm that is robust in the face of
both close encounters and highly eccentric orbits. We have also shown that
force-center switching during exceptionally close encounters
can be cleanly incorporated into the method and can substantially
enhance the stability of the algorithm without noticeably affecting its
desirable symplectic qualities. We have not, however, found a practical way to
regularize around the perturber while the switch is in effect; the stability
of this approach during highly eccentric {\it encounters} is correspondingly
questionable.

Our examination of Stark-based integrators indicates that they, too, are
subject to the radial orbit instability unless regularized, although
it tends to be less severe since the Stark approximation becomes
systematically better near the origin. Unless the perturbing potential is
{\it very} well represented by a Stark potential, they also appear
uncompetitive in terms of efficiency---by over an order of magnitude---due
to the cost of Stark steps relative to that of Kepler steps.
Among the Stark-based methods,
the regularized, symplectic method (\S~\ref{sec_rss}) consistently
outperformed the time-reversible method (\S~\ref{sec_trs}), in part because of
the linearly growing energy error exhibited by the latter.
Our conclusion is that Stark-based schemes are of marginal utility in the
integration of N-body systems.

It is clear that integrators based on a two fixed point (TFP)
splitting instead of a Stark or Kepler
splitting are also possible; they can be constructed in the same manner as the
Stark-based methods were. Such methods could be useful whenever two bodies
strongly dominate the mass in the system (e.g., asteroid motion in the
Sun-Jupiter system). As for Stark motion, however,
the relative expense of advancing the TFP Hamiltonian is a significant
handicap, and the circumstances in which its use is justified remain unclear.
On the other hand, since Stark motion is a subset of TFP motion it is not
unlikely that methods based on the latter splitting will generically
outperform those of the former type, since their analytic solutions are of
similar complexity.
It would be interesting to investigate this possibility in greater detail.

Although we have confined attention to the perturbed two-body problem, 
the techniques employed in this paper are also applicable to general
hierarchical $N$-body systems. In particular, we believe that
regularization of the $N$-body version of the
potential-splitting method (\cite{dunll97}) is likely to cure the instability
at high eccentricities noted by the authors. In principle force-center
switching of the kind described in \S~\ref{sec_ps} can also be done, but we
have not studied this possibility in detail.  In any event, we have found
the combination of regularization and potential-splitting
to be a powerful one, and to produce a remarkably versatile
symplectic method for the integration of nearly-Keplerian systems.

\acknowledgements

We thank Doug Hamilton, Norm Murray, Scott Tremaine, and Man Hoi Lee for
illuminating discussions.


\begin{thebibliography}{}

\bibitem[Abramowitz \& Stegun 1968]{abrs68}
Abramowitz, M., \& Stegun, I. A. 1968, Handbook of Mathematical
Functions (New York: Dover)

\bibitem[Bennetin, Galgani, \& Strelcyn 1976]{bengs76}
Bennetin, G., Galgani, L., \& Strelcyn, J. 1976, \pra, 14, 2238

\bibitem[Brouwer \& Clemence 1961]{broc61}
Brouwer, D., \& Clemence, G. M. 1961, Methods of Celestial Mechanics (New
York: Academic Press)

\bibitem[Danby 1992]{dan92}
Danby, J. M. A. 1992, Fundamentals of Celestial Mechanics (Richmond:
Willmann-Bell), p. 162

\bibitem[Dankowicz 1994]{dan94}
Dankowicz, H. 1994, \cmda, 58, 353

\bibitem[Duncan, Levison, \& Lee 1997]{dunll97}
Duncan, M. J., Levison, H. F., \& Lee, M. H. 1997, preprint

\bibitem[Gladman, Duncan, \& Candy 1991]{gladc91}
Gladman, B., Duncan, M., \& Candy, J. 1991, \cmda, 52, 221

\bibitem[Gladman et al.~1997]{glaet97}
Gladman, B. J., Migliorini, F., Morbidelli, A., Zappala, V., Michel,
P., Cellino, A., Froeschl\'e, Levison, H., Bailey, M., \& Duncan,
M. 1997, Science, 277, 197.

\bibitem[Grazier 1997]{gra97}
Grazier, K. R. 1997, Ph.D. dissertation, Dept. of Geophysics and Space Physics
(UCLA)

\bibitem[Hamilton 1992]{ham92}
Hamilton, D. P., \& Burns, J. A. 1992, \icar, 96, 43

\bibitem[Kinoshita, Yoshida, \& Nakai 1991]{kinyn91}
Kinoshita, H., Yoshida, H., \& Nakai, H. 1991, \cmda, 50, 59

\bibitem[Kirchgraber 1971]{kir71}
Kirchgraber, U. 1971, \cmech, 4, 340

\bibitem[Lee et al.~1997]{leeet97}
Lee, M. H., Duncan, M. J., \& Levison, H. F. 1997, in Computational
Astrophysics, Proc. 12th Kingston Meeting, ed. D. A. Clarke \& M. J. West (San
Francisco: ASP), p. 32

\bibitem[Lessnick 1996]{les97}
Lessnick, M. K. 1996, Ph.D. dissertation, Dept. of Mathematics (UCLA)

\bibitem[Levison \& Duncan 1994]{levd94}
Levison, H. F., \& Duncan, M. J. 1994, \icar, 108, 18

\bibitem[Levison \& Duncan 1997]{levd97}
Levison, H. F., \& Duncan, M. J. 1997, \icar, 127, 13

\bibitem[Marsden, Patrick, \& Shadwick 1996]{marps96}
Marsden, J. E., Patrick, G. W., \& Shadwick, W. F. (eds.) 1996, Integration
Algorithms and Classical Mechanics, Fields Institute Communications, Vol. 10

\bibitem[Mignard 1982]{mig82}
Mignard, F. 1982, \icar, 49, 347

\bibitem[Mikkola 1997]{mik97}
Mikkola, S. 1997, \cmda, 67, 145

\bibitem[Morbidelli \& Moons 1995]{morm95}
Morbidelli, A., \& Moons, M. 1995, \icar, 115, 60

\bibitem[Newman et al.~1997]{newet97}
Newman, W. I., Grazier, K. R., Varadi, F., \& Kaula, W. M. 1997, \baas, 29,
1102

\bibitem[Pars 1965]{par65}
Pars, L. A. 1965, A Treatise on Analytical Dynamics (London: Heinemann), p.
309

\bibitem[Rauch \& Ingalls 1997]{raui97}
Rauch, K. P., \& Ingalls, B. 1997, \mnras, submitted

\bibitem[Skeel \& Gear 1992]{skeg92}
Skeel, R. D., \& Gear, C. W. 1992, Physica, D60, 311

\bibitem[Skeel \& Biesiadecki 1994]{skeb94}
Skeel, R. D., \& Biesiadecki, J. J. 1994, \annm, 1, 191

\bibitem[Stiefel \& Scheifele 1971]{stis71}
Stiefel, E. L., \& Scheifele, G. 1971, Linear and Regular Celestial Mechanics
(Berlin: Springer-Verlag)

\bibitem[Wisdom \& Holman 1991]{wish91}
Wisdom, J., \& Holman, M. 1991, \aj, 102, 1520

\bibitem[Wisdom \& Holman 1992]{wish92}
Wisdom, J., \& Holman, M. 1992, \aj, 104, 2022

\bibitem[Wisdom, Holman, \& Touma 1996]{wisht96}
Wisdom, J., Holman, M., \& Touma, J. 1996, Fields Institute Communications,
10, 217

\end{thebibliography}
\end{document}